# Substrate-Versatile Direct-Write Printing of Carbon Nanotube-Based Flexible Conductors, Circuits, and Sensors

Crystal E. Owens[a,1], Robert J. Headrick[b], Steven M. Williams[b], Amanda J. Fike[a], Matteo Pasquali[b], Gareth H. McKinley[a,*], A. John Hart[a,*]

[a]*Department of Mechanical Engineering, Massachusetts Institute of Technology; Massachusetts, USA*
[b]*Department of Chemistry, Department of Chemical and Biomolecular Engineering, The Smalley-Curl Institute, The Carbon Hub, Rice University; Texas, USA*

Abstract

Printed electronics rely on the deposition of conductive liquid inks, typically onto polymeric or paper substrates. Among available conductive fillers for use in electronic inks, carbon nanotubes (CNTs) have high conductivity, low density, processability at low temperatures, and intrinsic mechanical flexibility. However, the electrical conductivity of printed CNT structures has been limited by CNT quality and concentration, and by the need for nonconductive modifiers to make the ink stable and extrudable. This study introduces a polymer-free, printable aqueous CNT ink, and presents the relationships between printing resolution, ink rheology, and ink-substrate interactions. A model is constructed to predict printed feature sizes on impermeable substrates based on Wenzel wetting. Printed lines have conductivity up to 10,000 S/m. The lines are flexible, with < 5% change in DC resistance after 1,000 bending cycles, and <3% change in DC resistance with a bending radius down to 1 mm. Demonstrations focus on (i) conformality, via printing CNTs onto stickers that can be applied to curved surfaces, (ii) interactivity using a CNT-based button printed onto folded paper structure, and (iii) capacitive sensing of liquid wicking into the substrate itself. Facile integration of surface mount components on printed circuits is enabled by the intrinsic adhesion of the wet ink.

Keywords

• printing • carbon nanotubes • flexible electronics • percolation • porous substrates

[1] Corresponding authors
*Email addresses:* crystalo@mit.edu (Crystal E. Owens), gareth@mit.edu (Gareth H. McKinley), ajhart@mit.edu (A. John Hart)

I. Introduction

Materials that combine mechanical strength, flexibility, and electrical conductivity are key to advances in flexible electronics, robotics, and medical devices. Mainstream approaches to scalable fabrication of electronics in non-wafer formats, such as for flexible displays, smart/RFID labels, and active clothing involve printing of liquid inks with conductive fillers [1]. Typical ink fillers include metal nanoparticles and nanowires, carbon nanotubes (CNTs), graphene (or graphene oxide) sheets, and conductive polymers [2]. The inks are selected for applications based on requirements for conductivity, flexibility, optical transparency, durability, and substrate compatibility.

For these uses, CNTs have advantages including high conductivity, low density, processability at low temperatures, and intrinsic mechanical flexibility. However, high-concentration CNT dispersions are difficult to formulate due to the low dispersibility limit of pure CNTs in almost all liquids, and the need for stabilizers and rheological modifiers, such as polyvinyl alcohol, to create a printable ink. The use of these modifiers inherently limits conductivity and requires post-processing by methods such as rinsing with alcohols [2], acid washing [3], or hot pressing
[4].

CNTs in solution percolate rheologically, or reach a semi-dilute interaction state, at low concentrations, often promoted by depletion-causing surfactants [5]. The formation of this continuous, interacting CNT network influences solution viscosity, processability, and the electrical performance of the final product. High ink concentrations can even generate fluid yield stress behavior [6]. As such, inkjet printing of CNTs has been possible only using dilute CNT dispersions, requiring several layers to generate conductive features [7, 8]. Conversely, a yield stress is beneficial for extrusion-based printing [9–16] and screenprinting [17, 18], as well as embedded extrusion printing into a soft matrix [19]. The presence of the yield stress improves feature retention, resolution, and smoothness.

By comparison, printed metal nanoparticles typically have excellent conductivity, yet their low material yield strain (e.g., < 0.5% for gold, silver) restricts flexibility of printed metal features. Meanwhile, the use of thinner metal layers to improve flexibility reduces conductivity, mechanical strength, and fatigue resistance. Alternative approaches use more complex manufacturing of wrinkled or pre-folded structures to allow macroscopic flexibility beyond the intrinsic fracture limit of the metal [20–22]. Similarly, promising inks made from nanowires, typically silver, show improved flexibility, though the metals remain inherently soft [1]. Metal particles are also prone to oxidation and sedimentation within ink suspensions [2], requiring protective and stabilizing additives, which in turn must be removed after printing, e.g., by sintering at high temperatures. Sintering in turn restricts printing to thermally compatible substrates, and can cause cracking of the metal. For the most stable (i.e., non-oxidizing) metals such as silver and gold, material cost is also of concern.

Moreover, scalable manufacturing of printed electronics demands compatibility with flexible, low-cost substrates [2]. Polymeric films are widely used in roll-to-roll printing of electronics due to their good mechanical properties, and resistance to oxygen and water penetration [2]. Polyimide and polyethylene terephthalate are predominantly used when high temperatures are required during processing such as for sintering of metal inks. Circuitry on paper has also been widely developed [23], such as for microfluidic diagnostics [24] and electrokinetic particle transport [25], disposable RFID tags [26], actuators [27], and sensors for electrochemistry [28].

Paper is readily modified, and clay and other coatings are commonly used to control ink sorption and fixation, as well as to adjust paper gloss and brightness [29] and to promote adhesion of high-surface tension inks [30].

Here, we describe the production and use of an aqueous CNT-based ink in a direct-write extrusion technique, producing conductive, flexible traces on a variety of impermeable (polymer) and porous (paper) substrates. In Section II.I, we describe a series of aqueous CNT inks having CNT volume fractions up to 2.4% and exhibiting shear thinning, yield stress rheological behavior. In Section II.II we relate process parameters (deposition rate) to feature size and conductivity during extrusion-based printing. In Section II.III we demonstrate the use of process control and percolation behavior to generate CNT traces with conductivity spanning 7 orders of magnitude, including low-concentration traces with a linear density less than $10^{-3}$ mg CNT/m. In Section II.IV we demonstrate CNT printing onto six flexible substrates–glossy paper, hydrophobic paper, vinyl ethylene laminating film, Kapton film, a viscose filter, polyethylene terephthalate, and chromatography paper–and introduce a model to predict the width of features on the impermeable substrates based on the effective Wenzel contact angle. In Section II.V we demonstrate printing of 2D structures with an integrated light-emitting diode, using the ink itself for mechanical adhesion and electrical contact to the circuit. Also, we create an interactive CNT paper button that illuminates an LED, utilizing the low contact resistance of the printed CNTs. Last, we combine a printed CNT capacitive sensor and CNT circuitry on chromatography paper to measure wicking fluid within the substrate itself.

II. Results and Discussion

*II.I. Formulation and Rheology of Aqueous CNT Ink*

To begin this study, a variety of CNT inks were produced with volume fractions of 0.02% < $\varphi$ < 2.4% CNT stabilized by sodium deoxycholate and Dowfax surfactants, which are known to be highly effective at suspending non-bundled CNTs at relatively high concentrations [31, 32]. Ink preparation involved sonication of CNTs in the water-surfactant mixture, followed by repeated addition of CNTs to increase concentration. Dispersions were centrifuged to remove bundles and impurities. In some cases, CNT concentration was further increased by dialysis as described by Maillaud, et al. [33]. CNTs were used from multiple sources for different inks, but uniformly had high aspect ratios, $\Lambda = L_{CNT}/d_{CNT} \approx 500-4500$, as measured previously [34] or reported by the manufacturer. The ink consistency ranges from a liquid to a paste, depending on the concentration. More details can be found in Section IV. The rheology of the CNT inks is tailored not by using polymeric fillers, as has been done before to achieve printability [35], but by using higher concentration and higher aspect ratio CNTs to create a fluid with a measurable yield stress. The absence of polymeric fillers or other rheological modifiers enables as-printed features to have high conductivity, without requiring post-processing. In addition, we believe that the yield stress contributes to shelf stability of the inks; we have observed that the inks printed up to two years after initial preparation showed consistent properties after printing, and no phase separation (e.g., no water layer below the solution, or hardened CNT layer on top).

At all concentrations used here, the CNT ink is shear-thinning. Above a threshold CNT concentration of $\varphi \approx 0.5\%$, a yield stress develops and increases with concentration, as shown in Figure 1a,b, ranging from 0.05 to 150 Pa with a strong power-law dependence, $\sigma_y \propto \varphi^{4.0}$, likely due to the long length of the CNTs [36]. The presence of a yield stress in the CNT suspensions is attributed to the inter-tube interaction of CNTs and the formation of nematic liquid crystalline domains (as shown schematically in Figure 1c), which is a feature of concentrated CNT solutions in various solvents as well as CNT suspensions with dispersing agents. [37–41] This is correlated with higher CNT concentration and length, which are predictors of higher conductivity [37, 38]. To verify this hypothesis, inks were imaged using transmission polarized light microscopy in Figure 1d. The opacity of inks increases with CNT concentration, and at higher concentrations (i.e., $\varphi > 0.3\%$), the inks become visibly birefringent. An isotropic-tonematic transition can be expected near $\varphi = 3.34/\Lambda \approx 0.1-0.7\%$ using the Onsager transition for rigid rods [34]. By comparison to this percolation threshold, isotropic conductive fillers like carbon black typically percolate electrically at $3 - 15\%$wt solution, while they do not exhibit liquid crystalline behavior [42, 43].

Upon shearing deformation at large strains, $\gamma >\hat{\ } 40\%$, mimicking the flow condition of extrusion printing, the CNT inks yield (Figure 1e). All inks restructure quickly and demonstrate no aging, having the storage modulus, $G^0$, stable over time (Figure 1f). Videos taken using transmission polarized light microscopy show similar absence of long-time change, with no visible change in liquid crystalline structure (e.g., coarsening) over time after shearing (SI Sec S1.I and SI Video S1). Additional rheological measurements showing small amplitude oscillatory shear tests of our CNT inks, and loss modulus data for the large strain amplitude sweep, are included in SI section S1.II.

Ink formulation to achieve the yield stress imparted by liquid crystalline behavior is key to its suitability for direct-write extrusion printing. In extrusion-based printing processes, higher yield stresses enable feature retention at smaller scales and allow greater feature complexity [9–11, 13–15]. The measured attributes of our CNT inks are also beneficial for printing by allowing extrusion through small nozzles at practical, low pressures followed by rapid recovery of full strength of the ink, imparting robustness to the printed structures [15].

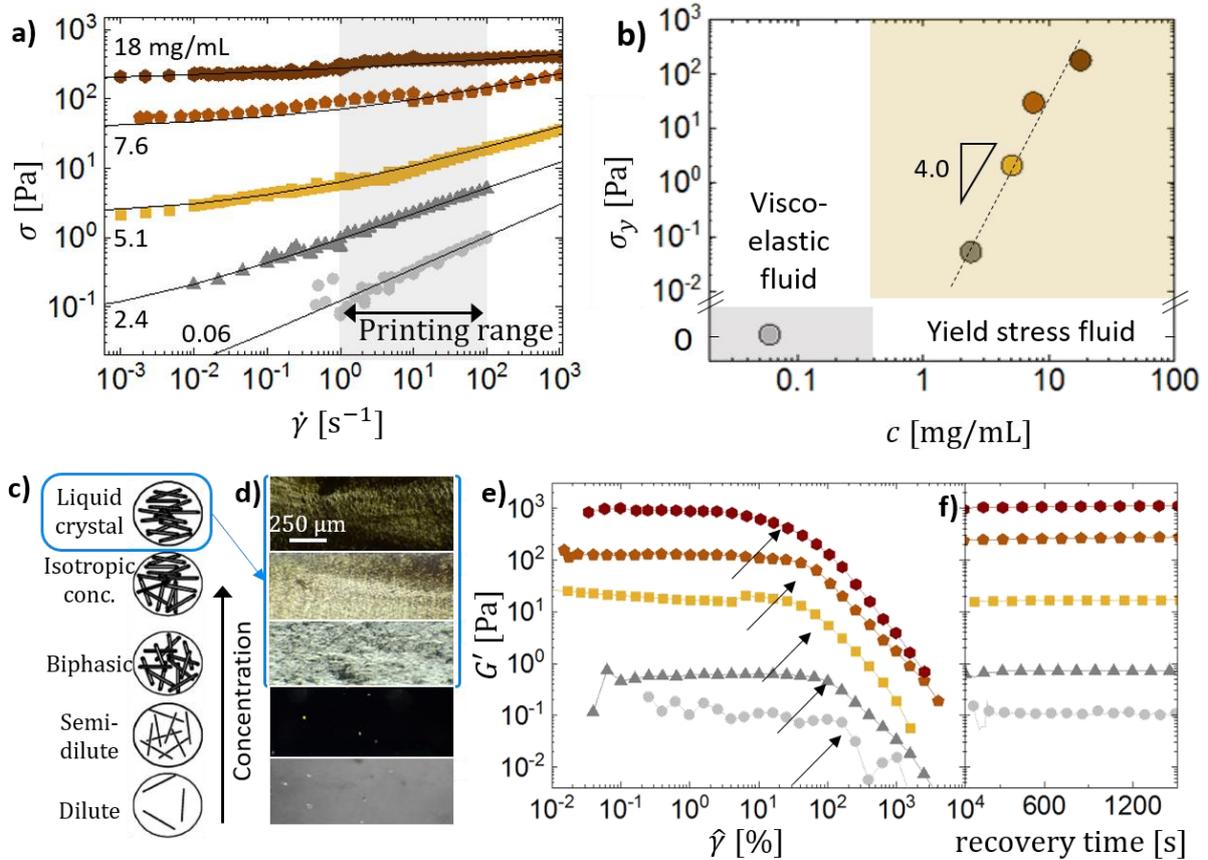

Figure 1: CNT ink formulation and rheology: (a) Flow curves of shear stress versus shear rate for CNT-based aqueous ink at five concentrations spanning $0.06 < c < 18$ mg CNT per mL or $0.02\% < \varphi < 2.4\%$ volume fraction, showing shear-thinning behavior and (b) a yield stress increasing sharply with CNT concentration. (c) As CNT concentration increases, CNTs eventually form nematic domains, which are (d) observed using polarized optical microscopy, shown here for five selected inks, in which the highest three concentrations exhibit birefringence. (e) As CNT inks are deformed under large strain amplitude, $\hat{\gamma} \gtrsim 30\%$, the elastic modulus collapses. (f) Upon cessation of shearing, the elastic modulus reaches a steady value within seconds and maintains a stable value over hundreds of seconds.

## II.II. Direct-Write Extrusion of CNT Ink

Direct-write extrusion of the CNT inks is performed using a modified 3D printer. Briefly, a custom syringe displacement extruder was constructed using a micrometer linear actuator (Thorlabs) and 3D printed components to hold the syringe. The syringe was affixed to the upper rail of a desktop printer (Makergear M2). The extruder enabled control of the ink volume dispense rate, ranging here from 0.2 μL/s to 9 μL/s. A stainless steel blunt-tipped needle (diameter = 0.12-0.60 mm) is used as the nozzle, and the ink is extruded with a small (0.1 mm) gap between the nozzle tip and the substrate, which in Fig 2 is a roughened polyimide/Kapton film (see Methods for more details). After extrusion, the solvent evaporates over 1-10 minutes, leaving solid CNTs.

During printing, ink is extruded using the 2D motion stage to control the nozzle in a digitally prescribed path, as depicted in Figure 2a-b. The concentration of CNTs in the printed features is independently controlled by adjusting the 2D nozzle motion speed, $v_{nozzle}$, and the average dispense speed of the ink, $v_{ink} = 4Q_{ink}/\pi D_0^2$. Here, $Q_{ink}$ is the volumetric flow rate of the ink through the nozzle and $D_0$ is the inner diameter of the nozzle. The nondimensional ink deposition rate, $v$, is therefore defined in Equation 1 as:

$$v = \frac{v_{ink}}{v_{nozzle}} \tag{1}$$

Similar process parameters have been used before to understand dimensional control in hydrogels, [44] while $v$ is defined here to best parametrize the influences of the printing process on geometry and electrical conductivity. A higher deposition rate increases the amount of ink deposited per linear distance, which also systematically increases the printed trace width, $w$, normalized as $w/D_0$. This scaling is examined by optical microscopy of printed features, and is plotted in Figure 2c,d. As the deposition rate increases, the resulting conductivity, $\sigma$, systematically increases as well. To adopt units commonly reported in the textile literature, we calculate the linear density of deposited CNT material in units of tex (mass per length along the axis of the printing trajectory), as in Equation (2).

$$\rho A_c = \frac{Qc}{v_{nozzle}} = cv\pi D_0^2/4 \tag{2}$$

Here, the mass density of printed CNTs is $\rho$ and $A_c$ is the cross-sectional area normal to the direction of printing. While neither property can be reliably measured at these size and mass scales, their product is readily calculated from input process parameters. We further note that we consider only the density and concentration from CNTs, not from residual surfactant. The initial mass concentration of CNTs in the liquid ink is $c$. The DC linear conductance, $\sigma A_c$, is defined as

$$\sigma A_c = L/R \tag{3}$$

where $R$ is the resistance measured by a four-terminal electrical probe, and $L$ is the length of this line trace that the resistance is measured across (see Section IV for details). In this case, linear conductivity, $\sigma A_c$, increases proportionally to linear density, $\rho A_c$ (Figure 2e) as would be expected for fully percolated (dense) conductive rod-like particles [45]. In addition, even at the lowest deposition rates, corresponding to the highest $v_{nozzle}$ and shear forces applied on the ink between the nozzle and substrate ($\dot{\gamma} > 100$ s$^{-1}$), alignment of CNTs is not observed. Thus, changes in conductivity are attributed to changes in linear density of the printed CNT traces alone (see SI Figure S3 and SI Section S1.III for details on CNT alignment).

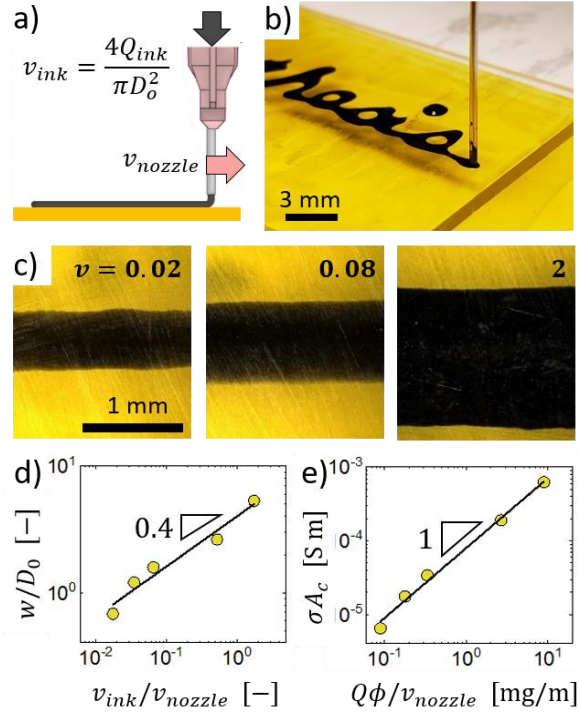

Figure 2: (a) CNT ink is dispensed at an average speed $v_{ink}$ from a nozzle of inner diameter $D_0$ moving at a speed $v_{nozzle}$ over a dry substrate. (b) Photographs of the experimental 3D printer writing cursive text onto a Kapton film. (c,d) As $v_{ink}/v_{nozzle}$ is increased, the resulting denser deposition of CNTs controllably increases both the dimensionless line width, $w/D_0$, and (e) the electrical conductivity of the printed line.

*II.III. Control of Conductivity During Printing*

After printing a CNT feature, the water takes up to 20 minutes to evaporate, depending on the total liquid volume and $v$, ultimately removing 95-98% of initial ink volume and leaving a layer of CNTs densified by capillary forces [46]. The conductivity of the printed, dry CNTs may be described by network percolation theory. While the high CNT aspect ratio contributes to a beneficial high yield stress in the liquid ink, it also enables formation of a conductive network at a low CNT loading [42], and thereby enables printed features to span a very wide range (~$10^7$-fold) of DC conductivity simply by control of the printing process parameters described above.

To show the percolation scaling, using inks with $\varphi$ = 0.02%, 0.7%, and 1%, straight lines were printed onto a porous chromatography paper and the conductivity was measured as shown in Figure 3a. By representing both density and conductivity as products of the cross-sectional area, $A_c$, percolation scaling laws can be applied to our data without need to measure $A_c$ (which is challenging on substrates, e.g., papers, with roughness greater than the thickness of the printed film). Percolation occurs at a linear density of $(\rho A_c)_{perc} \approx 1.0 \times 10^{-4}$ mg/m. Above this value, the conductivity ($\sigma A_c$) increases quadratically with the linear density ($\rho A_c$), in which the incorporation of additional CNTs increases both the connectivity and the density of the CNT network. At $\rho A_c \approx 1$ mg/m, the scaling transitioned to a linear relationship $\sigma A_c \propto (\rho A_c)^1$, indicating a change in electron transport, after which conductivity increases due to the increase of conductive mass alone. We observe that above this transition, the specific conductivity, $\sigma/\rho = \sigma A_c / \rho A_c$ is approximately constant. Finally, a lower bound for the conductivity of the CNTs in the network is estimated using the constant of proportionality between $\sigma A_c \propto \rho A_c$ in Figure 3a, predicting $\sigma_{CNT}$ & 3,000 S/m (see SI section S1.IV for full details). This indicates that there remains a large margin of possible improvement in conductivity up to the scale demonstrated for fully-dense and aligned CNT fibers, which is 10.9 MS/m, [47] and the expected conductivity of an individual CNT, which is around 100 MS/m. [48] However, the much lower conductivity of our printed CNT material is expected because of residual surfactant, the higher density of CNT-CNT contacts due to their lower alignment and lower surface area of contact, and the absence of doping, which cumulatively contribute to higher CNT-CNT contact resistance (compared to aligned fibers) as well as lower carrier density within CNTs [49–51] in the sparse isotropically-conductive network.

From these distinctive scaling laws, additional insights into the structure and nature of conductivity in printed CNTs can be made. First, the conductivity of the CNT network is dominated by bulk resistance rather than contact resistance [45], which is an indicator that the stabilizers in the CNT ink are not unduly preventing CNT network formation. Second, the conductivity indicates three-dimensional percolation, showing that the CNT network retains a three-dimensional, foam-like structure [42, 45]. This is likely enabled by the mechanical percolation within the ink resulting in a jammed microstructure and preventing full collapse. Third, as the data for our three inks overlaps, there is again no evidence of large-scale alignment, which would have increased the percolation threshold concentration, and offset the data sets accordingly [42]. These comparisons are also described in more detail in SI Section S1.IV.

Microscopic comparison of lines printed with three CNT inks at two deposition rates show key differences in morphology and quality (Figure 3b). Inks with higher CNT concentration and corresponding yield stress result in more narrow lines at equivalent $v$, as well as less wavy edges on the printed lines, as also shown in SI Figure S5. These features may be attributed to the fluid yield stress preventing spreading of printed lines, and to faster drying of more concentrated solutions. In addition, the concentration affects the ability to print continuous lines. At the lowest concentration, $\varphi = 0.02\%$ only, lines separate into segmented drops at low deposition rates. Higher concentrations of CNTs have been shown to stabilize fluid columns (i.e., via increasing extensional viscosity), slowing such breakup [52]; this allows printing of continuous lines, and may especially aid in high-speed printing of carbon nanotube inks.

We performed identical experiments and analysis using commercial CNT inks. For the same linear density of CNTs, our ink has, on average, 3-fold greater electrical conductivity and percolates at an intermediate concentration. The higher conductivity of our inks is likely due to the long CNT length in our inks, while the use of different surfactants, and CNTs with lower purity (i.e., higher amorphous carbon content) and quality (i.e., crystallinity) likely hinders the conductivity of commercial inks. These results are described more fully in SI Section S1.VI and SI Figure S6.

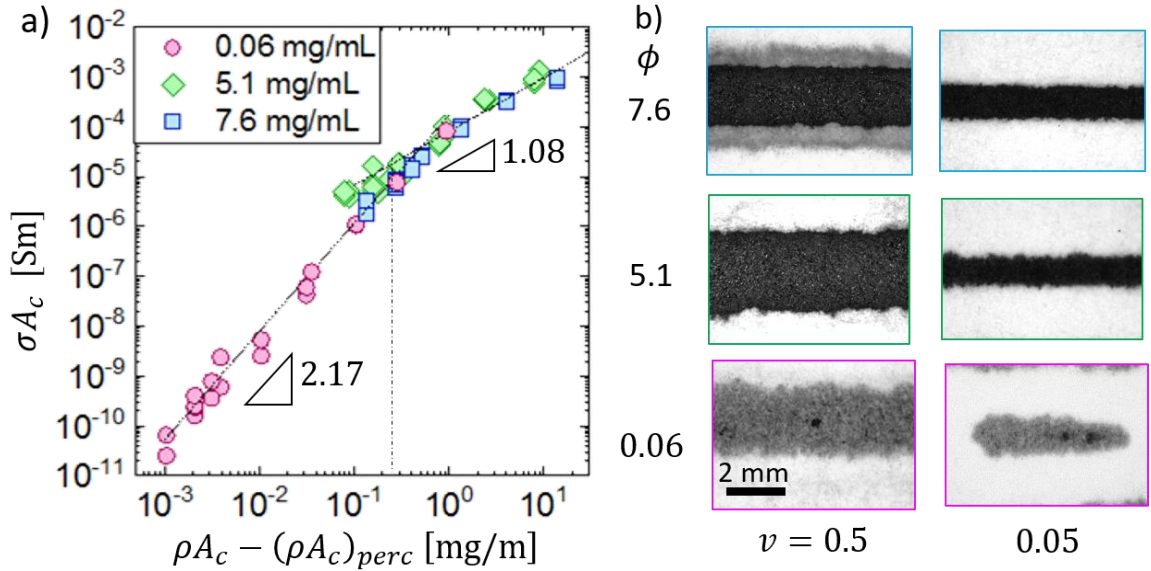

Figure 3: (a) The linear conductivity of printed CNT traces is controlled by the material deposited per area, and varies with a range of $10^8$ due to percolation effects. (b) Printed lines with CNT inks at different loading show enhanced opacity and increasing line edge smoothness with higher concentration of CNTs and higher $v_{nozzle}/v_{ink}$.

*II.IV. Ink-Substrate Interactions for Impermeable Substrates*

A wide variety of substrates are useful for printed and flexible electronics, and substrate selection depends on requirements for durability, abrasion resistance, and tolerance to temperature, humidity, and/or chemical exposure [2]. To understand the influence of the substrate on conductivity, feature size, and mechanical flexibility, we print CNTs onto a selection of substrates (Figure 4a-b): glossy paper, hydrophobic paper (glassine), Kapton film, vinyl ethylenecoated film, a viscose filter, polyethylene terephthalate (PET), and chromatography paper. The interaction between the ink and substrate is captured by the contact angle, $\theta$, which represents the balance of interfacial energies between liquid ink, solid substrate, and vapor from the ink at the three phase contact line. However, differing surface morphology and texture of the selected substrates has a confounding effect with the contact angle in governing the overall ink-substrate interaction. The texture is captured by the surface roughness, $r$, which is defined as the ratio of the true area of the surface solid compared to the normally projected area. In order to condense the menagerie of substrates to a single material parameter, we use Wenzel's law to assemble both terms into an apparent contact angle, $\theta^*$, where $\cos(\theta^*) = r \cos\theta$. This expression predicts that roughness amplifies the effect of native surface wetting properties.[53] In Figure 4, we show representative images of lines printed under identical conditions on each substrate including nozzle size (0.6 mm inner diameter), extrusion speed, and printing speed ($v = 0.5$). We also list values of $\cos(\theta^*)$ in increasing order from left to right, which transition from $\cos(\theta^*) < 1$ for impermeable substrates in Figure 4a to $\cos(\theta^*) > 1$ for porous paper in Figure 4b. When $\cos(\theta^*) > 1$, hemi-wicking [54] causes increased lateral flow along the paper, contributing to spreading that is ultimately limited by the deposited fluid volume rather than the contact angle.

Values of $r$ and $\theta$ for each substrate are tabulated in SI section S1.VIII.

From this concept of printing, a quasi-static equilibrium model was constructed to predict the width of lines based on the wetting parameter, $\cos(\theta^*)$, and the deposition rate, $v$, as depicted in Figure 4c and SI Figure S8. The Bond number, $Bo = \rho g h^2/\gamma$, evaluates the relative influence of gravity and capillary forces on maintaining the droplet shape, where $g$ is the acceleration due to gravity, $\rho$ is the mass density of the liquid ink, and $\gamma$ is the surface tension. When the deposition rate is small, surface tension causes a contraction of the line into a hemicylindrical trace, according to

$$\frac{w}{D_0 \tan\theta^*} = (\pi v)^{1/2}, \qquad (4)$$

which we refer to as the "2D droplet model". This is shown by a schematic inset in Figure 4c. When the deposition rate is large, gravity causes the fluid to spread laterally, as

$$\frac{w}{D_0 \tan\theta^*} = \frac{h}{D_0 \tan(\theta^*)^2} + \frac{\pi v D_0}{4h \tan(\theta^*)}, \qquad (5)$$

which we refer to as the "2D puddle model". Here, $h$ is set by the balance of gravity and surface tension, as

$$h = \sqrt{\frac{2\gamma(1-\cos\theta^*)}{\rho g}}. \qquad (6)$$

These models are compared to data in Figure 4c. Unreduced data is included in SI Figure S9. First, we observe that all data within our printable range is within the range of the droplet model due to the small volumes and small dimensions typical for printing. The droplet model adequately predicts the scaling of the printed line widths for all tested substrates and nozzle sizes, with best agreement for intermediate $0.05 < v < 1$. In the case of higher deposition rate, deviations from the model are attributed to more complex effects and solvent evaporation. Further details can be found in SI Section S1.VIII. We note that the model is generalized to capture the influence of printing our CNT ink onto different non-wicking substrates and to classify or rank these substrates. In addition, the model assumes no specific mechanism of ink application, nor specific details of the CNT ink rheology, but only the deposited volume of ink, the surface tension of the solvent, and the ink contact angle with the substrate. Given this baseline, it should apply more widely to printing of other liquid inks, and maskless printing methods such as inkjet printing that deposit similar relative volumes of ink. This would require further study to understand the influence of wetting across different printing processes.

For a dimensionless deposition rate of $0.5 < v < 50$ on impermeable substrates, the substrate influences the width of printed lines, but does not influence the conductivity (SI

Figure S4). To understand this more, we compared the morphology of CNT lines on different substrates. In Figure 4e, optical height maps show the surface texture of the hydrophobic paper and chromatography paper before and after printing CNTs. The bare substrates vary in structure and roughness, showing larger voids in the hydrophobic paper and extended cellulose fibers in the chromatography paper. CNT lines printed on both substrates have similar final texture, suggesting that the finer-scale morphology of the printed CNTs is insensitive to the substrate, and hence the conductivity is also insensitive to the substrate. This is similar to results obtained in screenprinting, for which printed layers beyond a critical thickness are no longer influenced by micron-level roughness of the substrate.[55]

As a further exploration, we studied the dynamics of the printing process from extrusion to the fully dried state by monitoring the resistance of the printed trace over time. This is particularly important because even bare CNTs are well-known to have humidity-sensitive resistance.[56–58] Our results showed approximately linear drying rates for each substrate, as is expected for small volumes,[59, 60] and drying times of 1-20 minutes for different substrates and volumes, as shown in SI Figure S10.

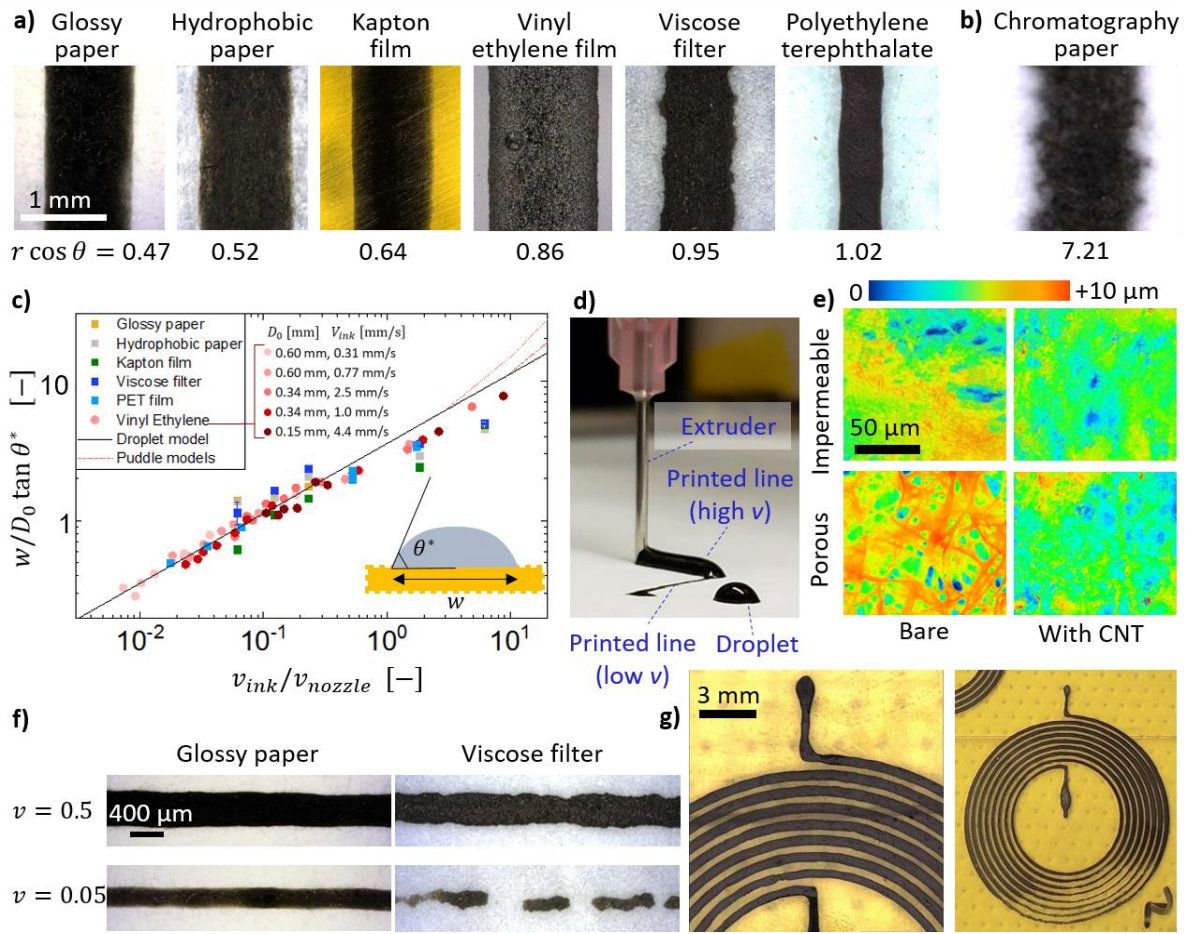

Figure 4: CNT ink printed onto several flexible substrates. (a) Here, we show traces deposited onto six inkimpermeable substrates with different wettabilities $\cos(\theta^*) = r\cos\theta$ and (b) one permeable chromatography paper. (c) The (reduced) line width of prints is consistent between substrates and print conditions, in agreement with an equilibrium wetting model balancing surface tension and gravity. The model cross section is shown schematically inset into the figure. (d) During the printing, liquid was initially large in volume for large $v$, later drying into a flat layer. When printing at small $v$, prints dried almost immediately, forming flat traces. (e) Topographic scans show the overall surface height variations ranging from dark blue (0 $\mu$m) to red (+10 $\mu$m), which are initially different on porous and impermeable bare substrates, but the surface of CNT traces printed on each results in a similar final texture. (f) The effect of wetting on different substrates is a limiting factor at minimum resolution, where the paper substrate, having lower $\cos(\theta^*)$, most readily held ink in place. (g) A spiral with narrow inter-line spacing is printed using $v = 0.07$ on Kapton film.

Though not analyzed in detail here, we also observed the influence of the substrate on feature fidelity. In particular, we observed that less rough substrates had better feature fidelity. This was evaluated by low edge tortuosity of printed features, defined as the perimeter length of the line edge normalized by the line length, as shown in SI Figure S12. In contrast, rougher substrates increase feature fidelity in inkjet printing when drop impact, contact angle hysteresis, and retraction of the fluid meniscus are also involved. [61]

In addition, de-wetting of ink from the substrate increases the variation of the width along a line at large scales, impacting feature fidelity, and ultimately limiting minimum feature size. The low wettability of viscose filter paper caused lines to become discontinuous at larger scales on viscose filter compared to the more strongly wetting glossy paper, limiting the minimum width of printed lines (Figure 4f, SI Figure S11a,b). Similarly, because line edges were smooth when printing onto Kapton film, printed designs could be spaced close together without shorting between them, for instance enabling printing of spiral patterns with a fine gap (see also SI Figure 4g).

*II.V. Application of Printed CNTs to Flexible Conductors, Circuits, and Sensors*

In the following sections, the fundamental ink formulation and printing process controls introduced above are applied to create functional artifacts on various substrates.

*II.V.1. Conformal and Extensible Circuit Elements*

Conductive traces are useful for delivering power to elements within a circuit, and flexible circuits must do this effectively during bending and twisting deformations. CNTs are known to be piezoresistive when subjected to in-plane strain, [62] and so here we focused on bending deformations, which generate lower material strains and can be applied to a larger variety of substrates. To this end, a CNT trace was printed onto a polyethylene terephthalate film with the name of one of our research laboratories, "mechano synthesis", using $v$ = 0.07. A gap between words held a blue surface-mount light-emitting diode (LED), which was manually placed onto wet ink immediately after printing so it adhered upon drying, as shown in Figure 5a-c. Using the relationships between material deposition rate and conductivity measured above, the words were designed to have a total resistance of 450 ± 15 Ω, allowing the 20 mA LED to be powered by a 9V battery.

We measured the relationship between the bending radius of the substrate and resistance for printed traces with linear density of 0.2-9 mg/m and radii of curvature from 0.02 mm (sharp fold) to 35 mm, corresponding to compressive strains of 1 to 0.002, respectively. Compared to the resistance of traces measured on a flat substrate (of $20 - 1200\Omega$, depending on the linear density), bending the substrate to a radius of $> 1mm$ show a change in resistance within 3% (Figure 5d). In addition, the conductivity of lines with the highest CNT linear density (9 mg/m) are less sensitive to bend radius even for lower bending radii. To demonstrate this consistent resistance, a sticker with printed CNT text and a LED was wrapped onto various objects including a glass vial and a fine rod, with radii of curvature of 35 and 1.5 mm, respectively (Figure 5eh). The LED remained illuminated, suggesting compatibility of printed CNTs to conformal electronics that can be printed in a planar configuration and then applied to 3D surfaces.

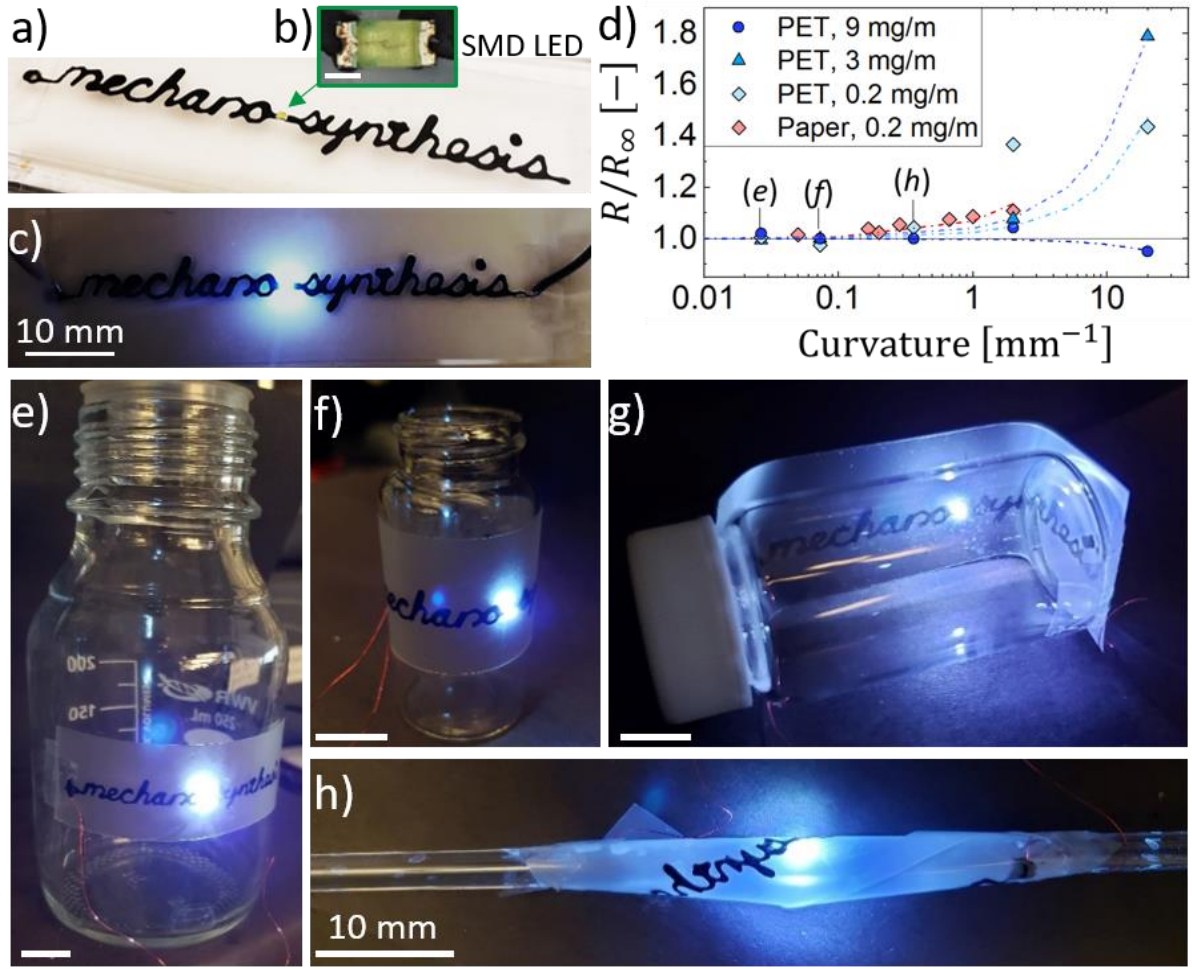

Figure 5: (a) A CNT trace was printed to spell "mechano synthesis" with (b) a blue surface mount diode (1 mm in length) manually placed between words (c) to illuminate when powered. (d) The conductivity of printed traces is constant within 3% when bent at a radius of >1 mm, which is shown by sustained illumination of the diode when wrapping the printed artifact around containers with radii of (e) 35 mm, (f) 13 mm, (g) 1.5 mm, and (h) 2.7 mm. Scale bars in (e-h) are all 10 mm.

*II.V.2. Interactive Touch Sensor*

Based on the percolation results of Section II.III, we judged that printed CNT networks have low contact resistance, which is useful for fabrication of electrical contacts. To understand this further, the contact resistance between printed CNT traces was measured to determine the dependence on CNT linear density and on applied pressure in a single cycle. As shown in Figure 6a, the contact resistance decreases with greater applied pressure for all paper substrates. For the more compressible chromatography paper, the change is more gradual than for the stiffer glossy paper. In addition, the contact resistance is proportionally larger for lower density CNT prints (Figure 6b. Here, CNT linear density was varied by the CNT volume fraction in the ink, $\varphi$, while the dimensionless ink deposition rate, $v$, was kept constant so the line width, and contact area, remained constant. Measurements are further described in SI Section S1.XII, along with contact behavior over several cycles of applied pressure, which shows no noticeable hysteresis.

Next, we built an interactive CNT-based button activated by contact between two printed CNT traces, inspired by children's books with interactive elements enabled by embedded electronics. The touch sensor is connected to activate an LED in the circuit, after the user presses the button. During printing, the LED was put in place using the printer nozzle to fill a small, 2 mm break in the printed line, which is shown in Figure 6c and SI Video S2. The CNT button uses an additional CNT line printed onto a folded section of paper. When digital (finger) pressure is applied to the button, it closes the circuit to illuminate the LED (Figure 6d-f). When pressure is removed, the folded paper relaxes, breaking the contact and opening the circuit. This button structure is used to create an interactive scene that was illuminated by pressing a location indicated on the page (Figure 6g).

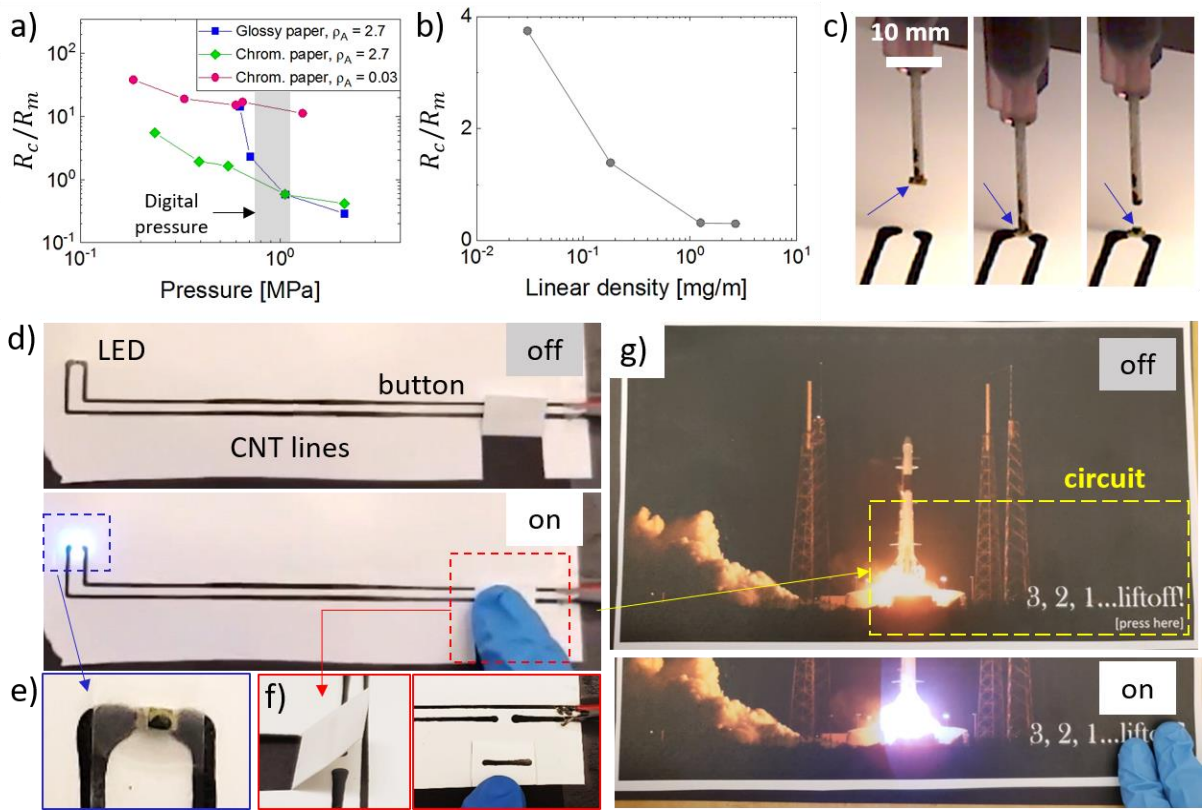

Figure 6: (a) Contact resistance is measured for CNTs printed onto papers as a function of applied pressure at the contact point. Here, contact resistance is normalized by $R_m$, the resistance of the CNT trace for 1 mm of length. (b) The contact resistance normalized by CNT network resistance decreases with printed linear density, indicating improved contact. (c) During printing onto glossy paper, the LED, indicated by the arrow, is placed by the printer at a designed 2 mm break in the printed line, making physical and electrical contact with the still-liquid ink. (d) The overall CNT circuit connected a power source, LED, and button that close the circuit to illuminate the power source. (e) The LED was protected with an adhesive overlayer to prevent detachment and (f) the button is made by a folded paper section holding a discontinuous line segment. (g) This blue LED is used to illuminated an image of a (SpaceX Dragon) rocket blasting off when the button was pressed through an overlaid illustrated page.

*II.V.3. A Capacitive Fluid Sensor*

Last, we demonstrate the ability to sense fluid motion, leveraging the printing of CNTs onto porous substrates that can imbibe liquid. As such, a device for sensing fluid wicking speed *in situ* or *in papyrus* was fabricated onto chromatography paper by printing a three-part device including (i) a CNT interdigitated capacitive sensor and with circuit elements to (ii) indicate the sensor is receiving power by lighting a red LED, and (iii) amplify the measured signal using a bipolar junction transistor (Figure 7a-c). Printing used $v = 0.07$ to maintain a high-quality printing range with well-prescribed line thickness; the printing process is shown in SI Video S3. The circuit is designed to change an output voltage signal in response to changing capacitance within the circuit, due to a changing dielectric constant of the spaces between CNT electrodes. The double-sided circuit allows use of a simplified design, and the front and back sides of the circuit are connected by vias constructed by pressing the dispensing needle through the paper. In a second design shown in Figure 7f, the capacitive sensor uses two parallel CNT lines that are on a separate sheet tethered by wire to the base circuit. While performance is similar, this allows for easier reuse of the circuit and ability to place it into smaller confined areas (like an Eppendorf tube).

For both circuits, the height of wicking fluid is measured by recording the sensor output and is verified by video analysis. The wicking followed Washburn's Law, in which the fluid height $h \propto \sqrt{t}$ for all tests. [63–65] After calibration, the fluid position and wicking speed are measured by the sensor within 12% (1.0 mm of height) for distilled water and isopropyl alcohol. The printed circuits are reusable once dry.

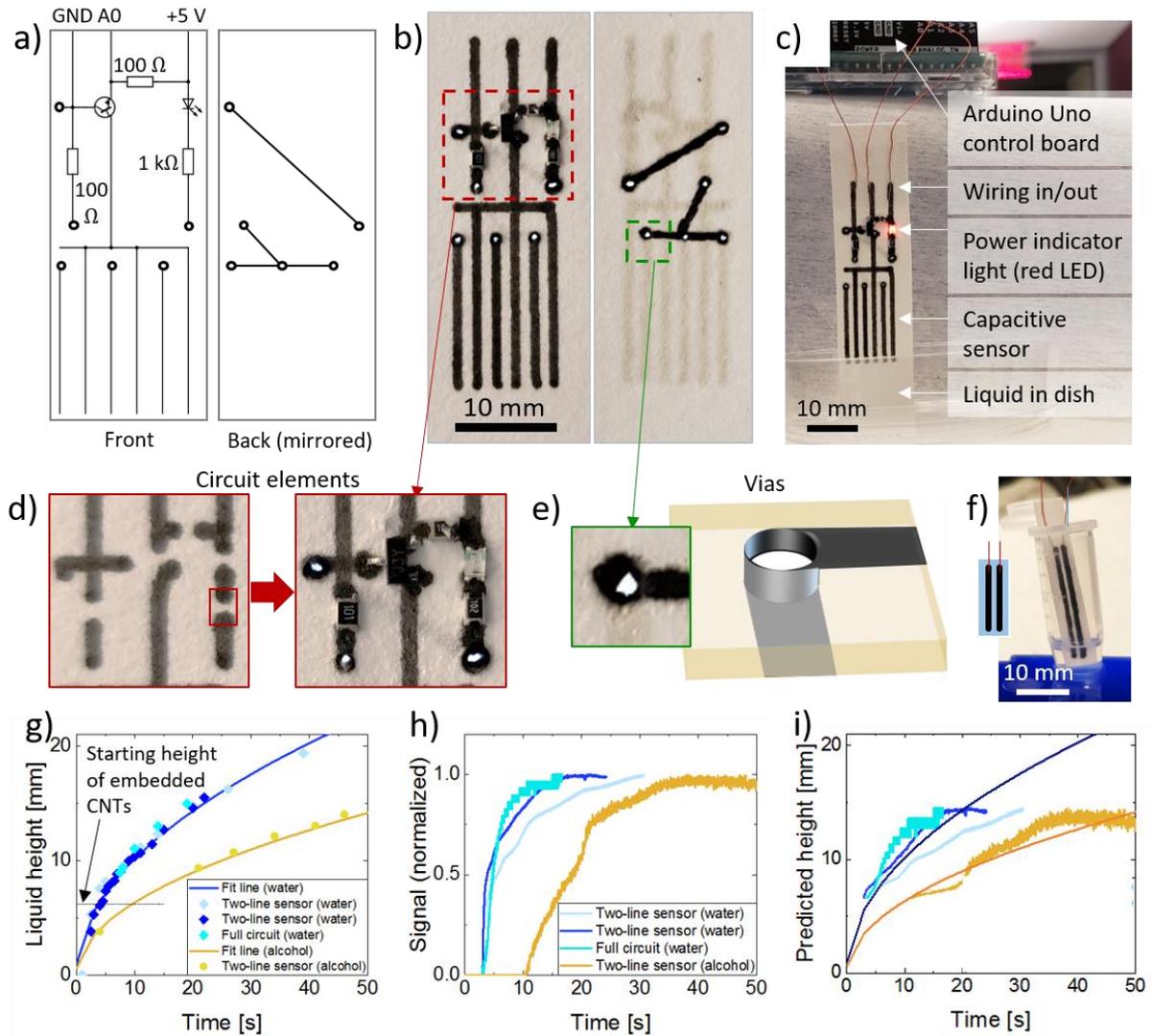

Figure 7: (a-c) A device for sensing fluid wicking speed is fabricated by printing a six-line interdigitated capacitive sensor with a two-sided circuit on a sheet of chromatography paper and (d) directly incorporating circuit elements to amplify the measured signal. (e) The two faces of the circuit are connected by conductive vias. (f) A second version of the circuit uses a two-line capacitive sensor connected by wires to the main circuit. (g) The height of wicking fluid is verified by video analysis, and (h) measured by sensor output. After calibration, (i) the wicking speed is measured within 12% (1 mm of height) for two different fluids (distilled water, isopropyl alcohol).

The capacitive sensor can also be used to interrogate droplet impact. We test this capability by adding droplets of distilled water of different volumes from 0.2 $\mu$L to 2 $\mu$L, which result in a sharp increase in capacitance within 50 ms of impact, followed by a steady decrease back to the baseline as the droplet spreads and dries (Figure S18a). Both the initial spike and the total area under the response curve vary systematically with the total volume over the entire tested range (Figure S18b). The sensor can also register addition of sequential droplets, and remains sensitive to the total volume of liquid deposited, whether that volume is added all together or as sequential droplets (Figure S18c,d). Such a sensor could be adapted to measure the influence of droplet impact on various porous substrates, such as face masks or granular beds.

*II.VI. Benchmarking the Conductivity and Flexibility Achieved by Direct-Write CNT Printing*

An Ashby-style plot, Figure 8, is used to compare the specific conductivity and the flexibility of the lines printed in this work to other materials. Available materials used for producing conductive traces include printable colloidal or nanoparticulate metals [66, 67], metal [68] and carbon nanomaterial [68, 69] foams, bulk carbon nanomaterial-based inks [10, 35, 70–72], and conductive polymers including PEDOT and PEDOT:PSS composites [73–78], filled thermoplastic polyurethanes [79], and zwitterionic polymers [30]. Also compared here are CNT-based fibers made by various spinning processes [80]. Here we only consider homogeneous conducting materials and CNT-based composites.

While the change in resistance with bending angle, and change in resistance after cyclic bending, presented here, are both typical measurements of flexibility and durability, the fine details of the measurements (e.g., bending radius, thickness of substrates and printed features, adhesion) are highly nonstandard and rarely reported in prior literature. In order to compare the flexibility of our printed ink with a spectrum of other printable inks, we compare the elastic compliance, or the inverse of the Young's modulus, $D = 1/E$. This compliance, in addition to the thickness of printed wires and substrates, adhesion, and printed geometry, strongly influence the mechanical behavior of flexible printed electronics. A benchmark flexibility is denoted as that of the substrate. When the compliance, $D$, of a print is too small/rigid, $D_{print} < D_{substrate}$, the printed material will stiffen the entire printed artifact and suffer the brunt of applied mechanical loads, reducing compliance and leading to failures of substrate adhesion.

Here, our printed CNTs exhibit an attractive combination of conductivity and compliance. The specific conductivity and compliance of several printed/printable materials is compared in Figure 8. Vertical bars indicate compliance of two typical substrates for printed electronics: Kapton [81, 82] and paper [83]. Bulk metals and metal features obtained by printing typically have excellent conductivity but low compliance compared to the substrates, while conductive polymers are generally extremely compliant (and sometimes extensible as well) but have significantly lower conductivity.

In comparison to other materials, CNT traces have a high maximum specific conductivity of $\sigma/\rho \approx 1.4 \times 10^2$ Sm$^2$/kg, and have a larger elastic compliance than the typical flexible substrates. Our printed CNT lines have a tensile modulus of 1.4 ± 0.7 GPa, an ultimate tensile strength of 38 ± 19 MPa, and a solid yield strain of 4.2 ± 2.1%, and exhibit brittle failure. Details of the mechanical characterization are included in SI Section S1.VII and SI Figure S7, but, in brief, we used tensile measurements of printed CNT wires that had been carefully removed from non-porous substrates. Metal foams/aerogels have similar flexibility and slightly lower conductivity, while spun CNT fibers have similar conductivity and much lower compliance. Other printed CNTs are typically more rigid due to the use of polymeric additives such as poly(vinyl alcohol) and thermoplastics to enhance printability, with significant loss of conductivity. We note that much of the published literature data for many conductive printed inks fell below the lower bound of the plotted data, with printed CNT inks ranging down to $\sigma/\rho$ < $10^{-11}$ S m$^2$/kg. In addition, data was projected for low-density CNTs, for which mechanical properties were extrapolated from other measurements, increasing the range of possible compliance as CNT mass is reduced.

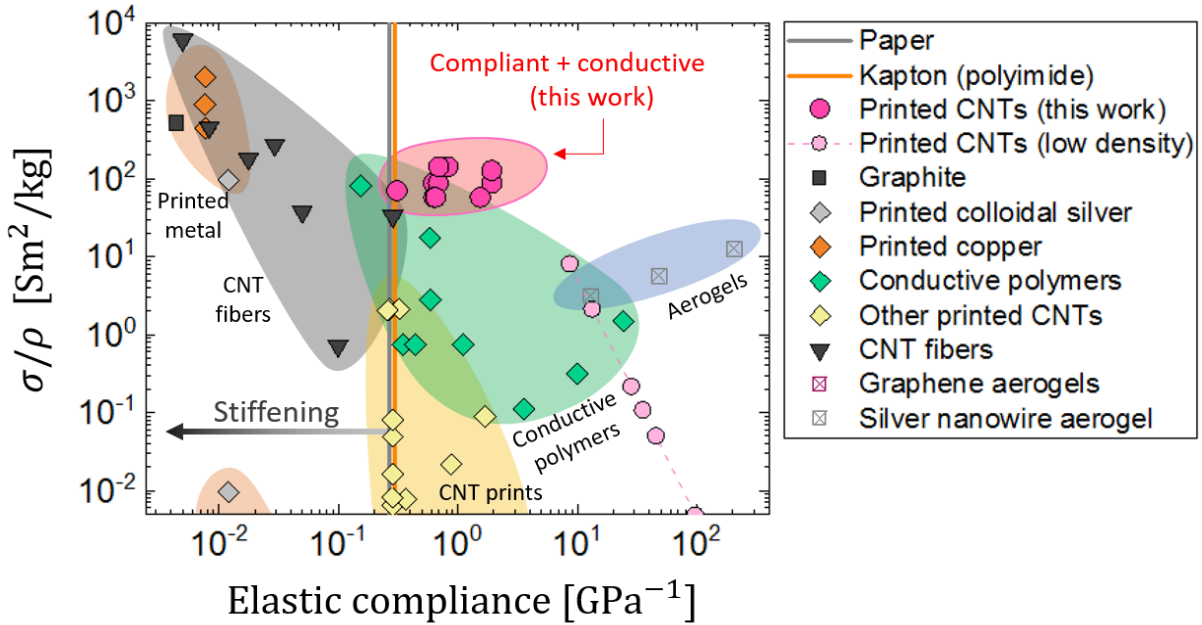

Figure 8: A comparison of the elastic compliance of conductive materials versus their specific conductivity. Below a compliance equal to the compliance of typical substrates (paper, Kapton), a material will increasingly stiffen the flexible electronic device. Above this level, the device is stiffened predominantly by the substrate, and we consider the printed material to be fully compliant. We targeted and achieved printed traces in the conductive and compliant material regime. We also include data for low-density CNTs (light pink circles) which have directly measured compliance, while the mechanical compliance shown here is projected from mechanical tests of the more dense printed CNTs, while the specific conductivity is directly measured for both.

CNT lines show stable resistance over repeated folding cycles, as shown in SI Figure S13a and in SI Section S1.XI. On both impermeable and permeable substrates, lines made with a linear density of $\rho A_c$ = 2.7 mg/m have a mean percent change in resistance under 2%. This behavior continues for sparser prints with $\rho A_c$ = 0.2 mg/m on impermeable substrates; however, resistance changes by over 30% for lines printed onto porous substrates with the more sparse linear density of $\rho A_c$ = 0.2 mg/m.

While the influence of bending to line conductivity is important for consistent performance of flexible electronics, extension is also relevant. CNTs have much larger strain-to-failure than metals, but are also known to be piezoresistive when large in-plane strain is applied. [62] Optimal stretchability requires not just a stretchable printed feature but also a substrate capable of extension, and often patterned substrate designs are used to most effectively allow macroscopic strain without damaging printed features, e.g. by using cleverly designed, non-affine deformations [20–22, 84].

Because minor flexion does not affect the printed CNT conductivity, we adapt the kirigami design work of Morikawa et al., [84] to design and create an ultrastretchable PET substrate capable of folding and macroscale extension, with our design and measurements shown in SI Figure S15. Using the equations presented by Morikawa et al., we tuned the simple kirigami pattern to employ only planar bending of the substrate, minimizing in-plane extension of the CNT network itself while achieving high macroscale strain. CNTs deposited on this substrate show DC conductivity with under 5% change for single cycle up to 830% (maximum) strain of the base substrate, or similarly under 5% change for 100 cycles up to 400% strain each, and under 1% change in resistance for 100 cycles at 200% strain. This extensibility can help one-time integration into devices (e.g. for thermoforming) and for repeated cycles of moderately large (<500%) strain.

III. Conclusions

We have described the development and application of a series of aqueous, CNT inks for use in direct-write printing of electronically functional features and devices. We showed in detail how the properties of the ink determine the electrical conductivity of printed features, and the related processing capabilities (feature size, production speed) required for manufacturing success. While the extrusion-based methods presented herein are best suited to localized deposition, the ink could readily be adapted to production-scale methods, such as gravure and screenprinting, that can readily handle and benefit from the rheological properties of yield-stress fluids. Moreover, the ability to process and deposit CNTs from aqueous inks can be beneficial to the development of low-cost and wearable sensors, RFID tags, disposables, and morphable structures.

IV. Experimental Methods

*Materials and Preparation of Carbon Nanotube-based Aqueous Ink:* Carbon nanotube (CNT) inks were prepared to concentrations between 2.4 mg/mL and 7.6 mg/mL of CNTs with sodium deoxycholate (DOC) surfactant to stabilize the suspensions with 2 wt% (20mg DOC / g solvent) for each suspension. The CNTs were tip sonicated for up to 10 hours in water (10 mL) with DOC (0.2 g), adding more CNTs over time (24-76 mg) until the desired concentration was reached. The CNTs used (EC1.5, Mejio, Japan for 2.4 and 5.1 mg/mL) have a mean diameter of 1.5 nm and initial length of 9.4 $\mu$m [34]. For the lowest concentration ink, 0.06 mg/mL, a low-concentration ink was prepared and then centrifuged for long times, keeping the supernatant. For the highest concentration inks, 7.6 and 18 mg/mL, Tuball (Coat E) (20 mL) was dialyzed in 5-10%wt Dowfax and dialysis tubing was placed in the reservoir and stirred for several days to concentrate the ink, using the method introduced by Maillaud et al. [33] The final CNT concentration in each ink was determined using the absorbance (550 nm) of serial dilutions of the CNT ink compared to standards of known concentration with the same surfactant concentration, using Beer's Law. Four commercial CNT inks using single-walled CNTs were printed "as-is" alongside our in-house created CNT inks for comparison: CG300 (SWeNT), Coat

E (OCSiAl), and Invisicon 3500 and 3400 (Nano-C). For the inks we prepared, CNT concentration, $c$, was converted into volume fraction, $\varphi$ by dividing by the CNT mass density, $\rho_{CNT}$, using

$$\rho_{CNT} = \frac{4000}{A_s(d+\delta_{vdW})^2}[nd - 2\delta_{vdW}\Sigma_{i=0}^{n-1}i] \tag{7}$$

where $A_s$ = 1315 m$^2$/g, $\delta_{vdW}$ =0.34 nm, $n$ is the number of walls of the CNT, and $d$ is the diameter, from [34].

*Rheological Measurements:* Rheological behavior of the inks was measured with a stresscontrolled shear rheometer (DHR-3, TA Instruments) using a cone-and-plate fixture (20 mm diameter, 4° cone angle; TA Instruments) with adhesive sandpaper having 10 $\mu$m roughness (Trizact A10; 3M) applied to prevent slip of the fluid sample. The shear viscosity was measured by decreasing the shear rate from the maximum to the minimum value to allow the sample to equilibrate. Small amplitude oscillatory shear tests were performed with strain control at $\hat{\gamma}$ = 1% strain. Strain amplitude tests were performed at $\omega$ = 1 rad/s, and the subsequent recovery test was performed at $\omega$ = 1 rad/s and $\hat{\gamma}$ = 10% strain to obtain a measurable signal while remaining within the regime of small (linear) deformation.

*2D Printing of CNT Inks:* A 3D printer (MakerGear M2, MakerGear) used a custom syringe displacement system based on a micrometer-resolution single-axis translation stage (PT1, Thorlabs) to print the CNT inks at room temperature. The printer used a 1 mL capacity syringe and sterile blunt-tipped metal needles (Sanants and Nordson). The orifice size of the needles ranged from 0.12 to 0.60 mm inner diameter (30 to 20 gauge). The lateral speed of the printhead ranged from 10 to 3,000 mm/min, and the extrusion flow rate ranged from 0.2 $\mu$L/s to 9 $\mu$L/s. After printing, samples were left to fully dry in ambient conditions for 20 minutes before further characterization. Substrates used here include regenerated cellulose (10410214, Whatman), chromatography paper (3MM, Whatman), glossy paper (Xerox digital color), hydrophobic paper (glassine, Cole Palmer), Kapton film (Myjor, 1 mil thickness), polyethylene terephthalate (IVict) and laminating film with heat-sensitive coating (TYH Supplies). Kapton film was roughened with sandpaper (P2000) before printing to promote ink adhesion.

*Characterization:* Measurements of printed line width were made using an optical microscope (Zeiss SmartZoom 5). Surface roughness was measured using a laser scanning confocal microscope (Keyence VK-X250) and averaged over an area for each substrate. The conductivity of CNT prints was measured using an LCR meter with 4 terminal Kelvin clips (IM3536; Hioki) at room temperature (23-25° C) and ambient humidity (23-52% RH) at least 20 minutes after printing or when lines were visibly dry. Scanning electron micrograph images were taken using a high-resolution SEM (Zeiss Merlin) on printed lines without sputtering or other modification. Linear density was calculated using process parameters, as described in Equation 2.

*Functional artifacts:* Integrated circuit elements included LEDs that were 20 mA surfacemount units from a reel (Chanzon 0603 SMD LEDs, 1.6 mm x 0.8 mm), a J3Y transistor, and standard 101 and 102 SMD resistors. These were attached manually or using the printer nozzle, using the same wet CNT ink for mechanical adhesion and electrical connection.

V. Supporting Information

Supporting Information is available from the Wiley Online Library or from the author.

VI. Acknowledgments

Financial support was provided by the NASA Space Technology Research Institute (STRI) for Ultra-Strong Composites by Computational Design (US-COMP, grant NNX17AJ32G), the U.S. Air Force Research Grant No. FA9550-15-1-0370, the Robert A. Welch Foundation Grant C-1668, and the Department of Energy (DOE) awards DE-EE0007865 (Office of Energy Efficiency and Renewable Energy and Advanced Manufacturing Office) and DE-AR0001015 (Advanced Research Projects Agency-Energy). C.E.O. was supported by the United States Department of Defense (DoD) through the National Defense Science Engineering Graduate Fellowship (NDSEG) Program. R.J.H. was supported by a NASA Space Technology Research Fellowship (NSTRF14), Grant No. NNX14AL71H. This work made use of the MRSEC Shared Experimental Facilities at MIT, supported by the National Science Foundation under award number DMR-1419807. The authors further acknowledge support from the Soft Matter Group within the Materials and Manufacturing Directorate of the Air Force Research Laboratory, and the authors gratefully thank Drs. Benji Maruyama, Chris Macosko, Anoop Rajappan, Megan Creighton, and Gaurav Chaudhary, and Olek Peraire for insightful discussion on different aspects of this project.

VII. Conflict of Interest

The authors declare no conflicts of interest.

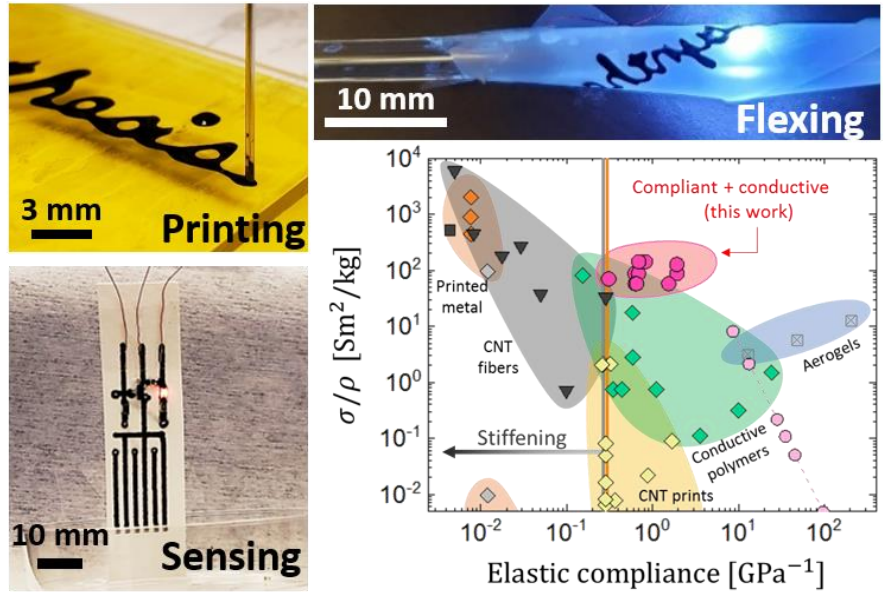

Table of Contents

A direct-write printing process is controlled by two parameters to extrude aqueous carbon nanotube (CNT) solutions with liquid crystalline behavior, drawing highly conductive and flexible traces, sensors, and circuits with integrated circuit elements onto a variety of flexible substrates. Solution quality and high CNT density allow high final conductivity.

S1. Supporting Information

*S1.I. Included Videos*

In SI Video S1, a time-lapse video shows two CNT inks (0.7%, 1% volume fraction) sitting statically in a glass capillary over a length of 15 minutes starting directly after shearing outside the microscope. These are imaged using transmission polarized microscopy, and show no detectable motion, coarsening, or relaxation after 2 minutes, in agreement with rheological recovery data shown in Figure 1f.

In SI Video S2, we show some steps in printing of the picture book with interactive CNT touch sensor, including making CNT traces, placing the LED, and lighting the LED using the button on the circuit and with the cover picture.

In SI Video S3, we show some steps in printing and using the fluid sensor, including printing CNT traces, powering on the device with illumination of the power indicator LED, and addition of water.

*S1.II. Time-Dependent Rheology of CNT inks*

Following from Figure 1, we show the remaining Herschel-Bulkley fit parameters, $k$ and $n$, as a function of CNT volume fraction, $\varphi$ in SI Figure S1. We also include the frequency response of the inks in small amplitude oscillatory shear tests in SI Figure S2a, and the storage and loss modulus response of two inks to high amplitude deformation, and subsequent recovery in SI Figure S2b,c.

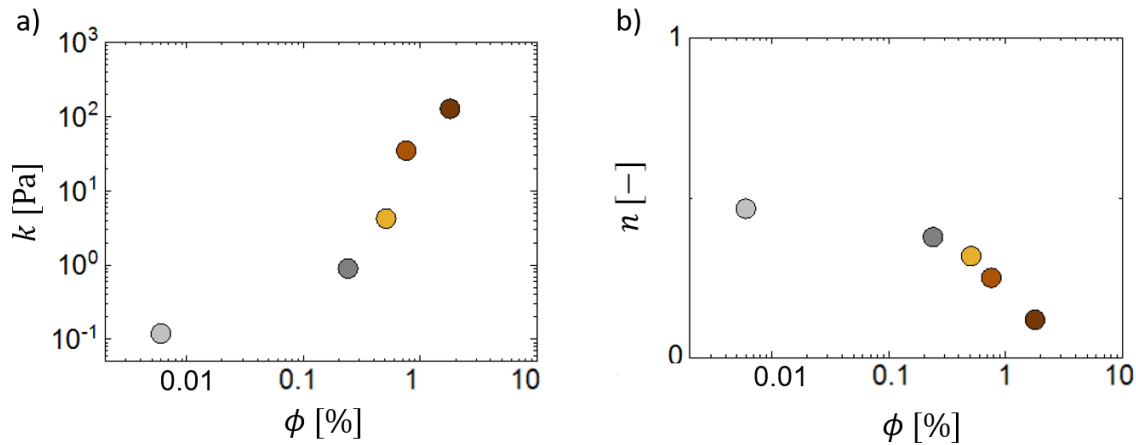

Figure S1: Evolution of (a) the consistency index, $k$, and (b) shear thinning index, $n$, for the inks shown in Figure 1 as a function of CNT volume fraction, $\varphi$.

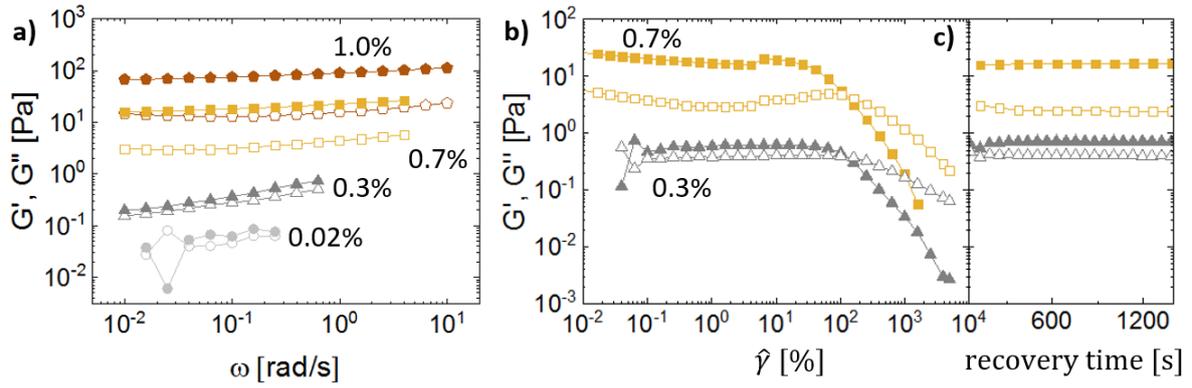

Figure S2: Following from Figure 1, we show (a) the frequency response of the inks in small amplitude oscillatory shear tests and (b) the onset of nonlinearity at large strain amplitude and subsequent recovery. Storage moduli, $G'$, are shown by filled symbols while loss moduli, $G''$, are hollow symbols with the same color and shading.

*S1.III. Evidence of Isotropic CNT Arrangement*

High shear forces have been shown to cause alignment of CNTs and other elongated nanostructures [9, 85]. In our printing process, we imaged printed lines on chromatography paper using transmission polarized microscopy and found no evidence of large-scale alignment induced by printing, as shown in SI Figure S3, which may be due in part to competing elastic instabilities observed in shear flow of CNTs [86, 87]. Additional scanning electron micrographs and polarized Raman scans also showed no orientation, though these null results are not included here.

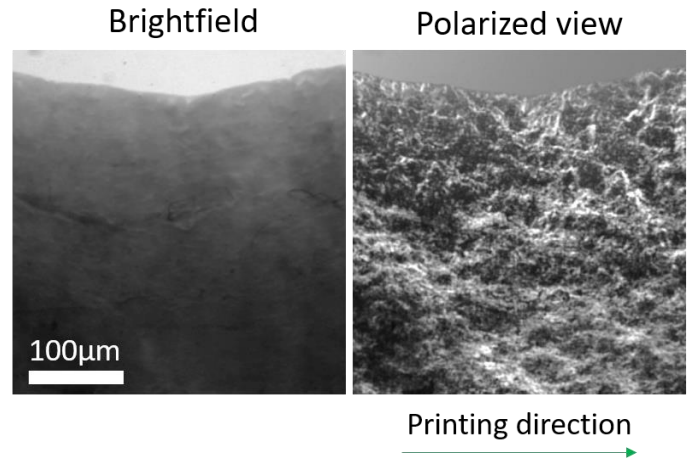

Figure S3: The edge of a printed CNT line was imaged under transmission polarized microscopy, showing in brightfield and polarized light. No long-range order is apparent.

*S1.IV. Calculating Conductivity of Constituent CNTs*

The conductivity of a percolating network of CNTs is described well by the typical percolation law in Equation (S1) [42, 88]

$$\sigma_{system} = \sigma_{CNT} f(\varphi - \varphi_c)^t \tag{S1}$$

where $\sigma_{CNT}$ is the bulk conductivity of the constituent CNTs, $t = 1$ is the exponent in the linear regime past percolation, and $\varphi$ is the volume fraction. For an aspect ratio of $\Lambda = L_{CNT}/d_{CNT}$, the percolation volume fraction is $\varphi_c \approx 2/\Lambda \approx 0.04\%$, and an empirical dependence on the aspect ratio of CNTs is [88]

$$f = 10^{0.85[\log(\Lambda)-1]} \tag{S2}$$

where $f \approx 50$ when we use the aspect ratio predicted from the observed transition concentration to liquid crystalline behavior in Section II.I ($\Lambda \approx 1000$). Using a theoretical scaling [89], which we rearrange and rewrite here for convenience,

$$f = \Lambda^{0.087\log(\Lambda)} \tag{S3}$$

giving $f \approx 6.1$. We convert our calculated CNT mass density to the volumetric fraction of CNTs in air by

$$\varphi = \rho/\rho_s \tag{S4}$$

where $\rho_s$ is the bulk density of CNTs (1.2 g/cm³), and rewrite equation S1 in terms of known quantities during printing to calculate the bulk conductivity of CNTs comprising our printed lines for $\phi \gg \phi_c$ as

$$\sigma_{CNT} = \frac{\sigma A_c}{\rho A_c} \frac{\rho_s}{f} \tag{S5}$$

which is found to be constant as in Figure S4b giving $\sigma_{CNT} \geq 5{,}000$ S/m using the average of the $f$-values calculated above. This conductivity is expected to be a minimum bound for the bulk conductivity of the CNTs, and is reduced from the true value of pure CNT conductivity due to contact resistance and effects from network topology [42, 89].

In comparison, the conductivity of the liquid inks before solvent evaporation was 0.17 ± 0.15 S/m and 1.7 ± 1.5 mS m²/kg, corresponding to a much lower CNT network conductivity of $\sigma \approx 27 \pm 35$ S/m, indicating significantly less CNT-CNT contact in solution than after ink has been deposited. This is to be expected from a stabilized suspension.

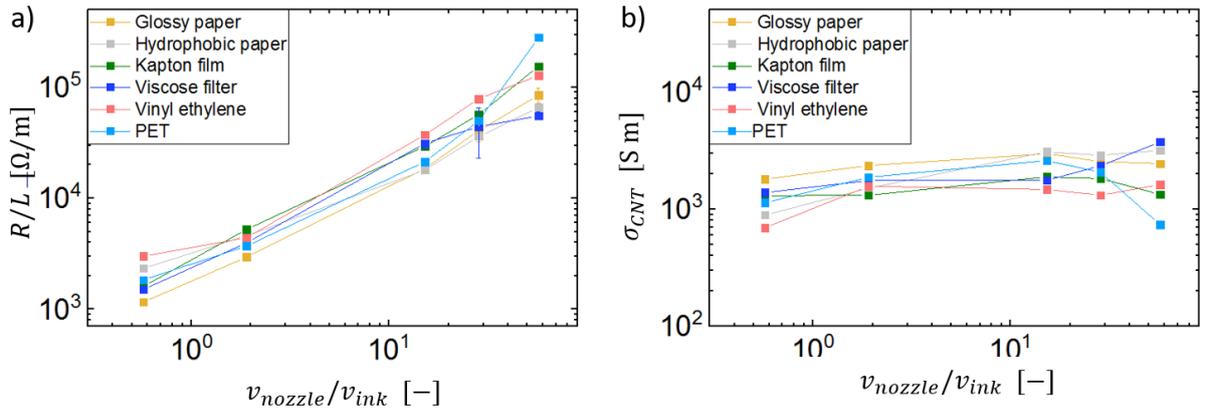

Figure S4: (a) For CNT lines printed onto a range of substrates, the conductivity increased linearly with deposition rate. (b) Using equation S5, the conductivity of the constituent CNTs, reduced by the contact resistance of the network and residual surfactant, was estimated to be 1,900 S/m.

We observe scaling laws of $t = 2$ directly after percolation and $t = 1$ after a further transition, as shown in Figure 3 of the main manuscript. Immediately after percolation, a two-dimensional system would behave as $t = 1.33$ in equation S1 and a three-dimensional system would be expected to show $t = 2$ [42], indicating that the CNT networks in the present work form a three-dimensional network.

When conductivity of a 3D network of elongated rods is dominated by contact resistance between individual rods, the scaling exponent should be $t = 2$ after percolation, without a transition region, due to charge mobility being limited by the number of contacts between individual rods throughout the network. [88] However, when contact resistance is relatively low, the network resistance is dominated instead by bulk resistivity of CNTs. Then, the scaling law of $t = 2$ is expected, where the overall number of contact points is still limiting, with a transition to $t = 1$ when the number of contact points has saturated. [88] When the contact resistance and bulk resistance are comparable, this exponent is expected to vary between the lower and upper bounds depending on the precise ratio and the material aspect ratio [90].

A flow-induced or shear-induced alignment of anisotropic materials increases the percolation threshold required for onset of conductivity, which would shift the measured percolation curve laterally [42].

*S1.V. Influence of Concentration on Feature Size*

Lines were printed onto paper using CNT-based aqueous inks with three different concentrations of CNTs, and the widths of the lines, $w$, were measured after printing and compared to the extruder nozzle inner diameter, $D_0$. As shown in SI Figure S5, as the CNT concentration increased, the line width systematically decreased. This is attributed to a higher yield stress in the higher-concentration inks, as shown in Figure 1b.

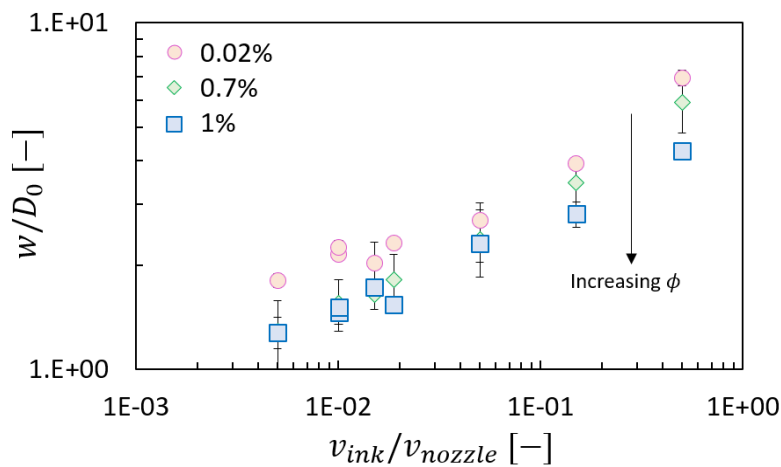

Figure S5: The concentration of CNT ink affected the width of printed lines. For a range of deposition rates, the low concentration ink systematically spread wider than the high-concentration inks, which is attributed to the yield stress rheology of the high concentration inks.

*S1.VI. Comparison of Commercial CNT Inks*

Four commercial CNT inks using CNTs in aqueous suspensions were printed "as-is" alongside our in-house CNT inks using the same process and control parameters. These included CG300 (SWeNT) with single-walled CNTs in water at 1.00 mg/mL of CNTs, Coat E (OCSiAl) with single-walled CNTs in water with sodium dodecylbenzene sulfonate surfactant around 4 mg/mL of CNTs, and Invisicon (Nano-C) with single-walled CNTs in IPA and water at approximately 5.2 mg/mL for Invisicon 3500 and 3.4 mg/mL for Invisicon 3400. All CNT concentrations are reported by the manufacturers. As shown in Figure S6, the behavior was similar for all inks, roughly following a quadratic relationship between linear density and linear conductivity. The commercial inks exhibited a mean conductivity of approximately 30% of the electrical conductivity of our inks within the testable range, with the highest differences in conductivity of 7% of the value of our inks at moderate CNT concentration, and a 5-fold greater conductivity at the lowest concentration of the CG300 ink, as both the CG300 and Coat E inks appear to percolate at lower thresholds than our ink, causing greater conductivity at low concentrations. Meanwhile, the average values of conductivity of CNT items reported in published papers is still lower, at most approximately 0.8% of the specific electrical conductivity of our inks (refer to Figure 8 for details), which may be attributed to the use of additional polymeric fillers in those papers.

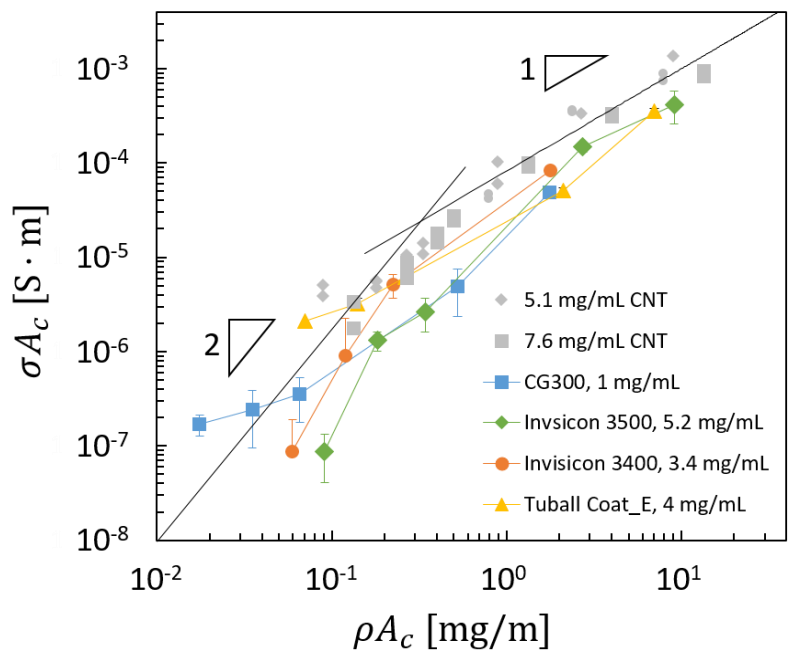

Figure S6: Three commercial CNT inks show similar behavior as CNT inks formulated in-house, denoted as grey symbols. Greater noise in the data for the commercial system is ascribed to use of lower concentrations. Error bars show standard deviation of 5 repeated tests.

*S1.VII. Mechanical Testing of CNT Prints*

CNT lines were printed onto and carefully removed from impermeable substrate with low adhesion to test the mechanical properties of the printed and dried CNT material (SI Figure S7a). Examination of the failure surfaces after a tensile test showed simple fractures, with no apparent stretching or CNT pull-out, agreeing with our previous determination of no apparent orientation within the CNT prints. Tensile data is presented in SI Figure S7c-f.

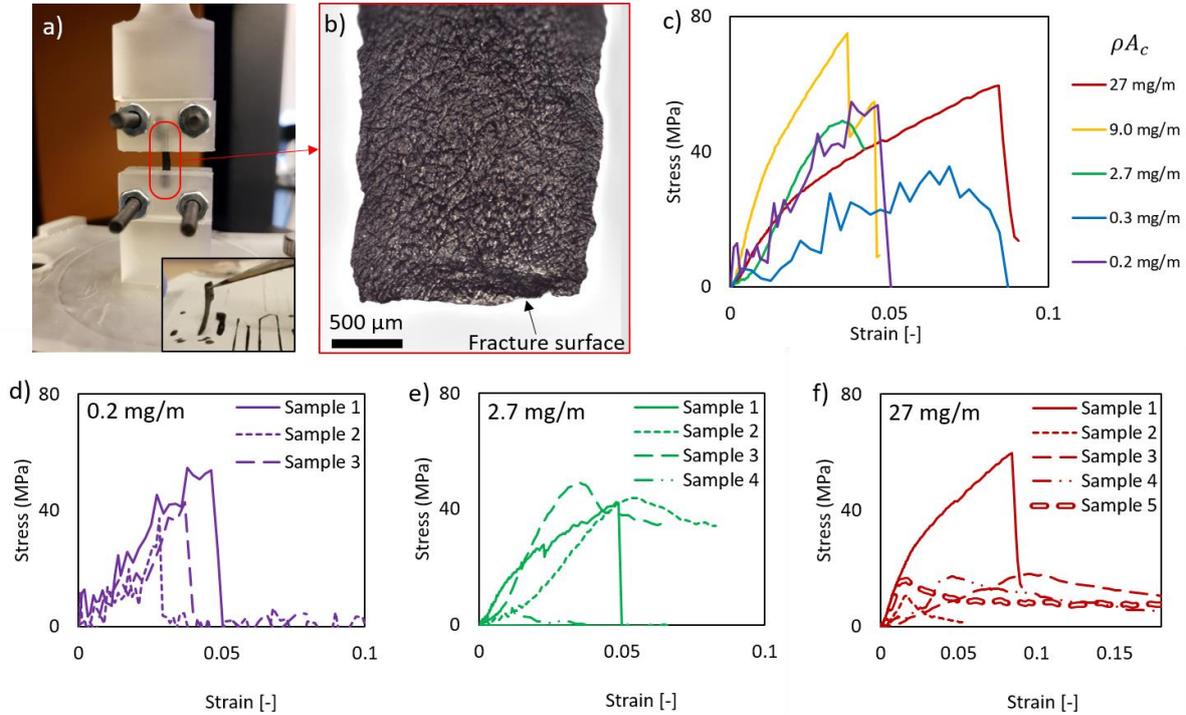

Figure S7: (a) CNTs were printed, dried, and removed from the low-adhesion substrate to directly test their mechanical properties. (b) The typical failure surface was a simple fracture, with no apparent CNT pull-out. (c) Tensile data is presented for lines with linear density from 0.2 to 27 mg/m, and (d-f) repeated tests are shown for three selected linear densities, showing similar elastic moduli among all tests, while the failure strain and strength were more variable, particularly for the largest sample.

*S1.VIII. Predictive Model for Width of Printed Lines*

A model was constructed to describe the width of lines as printed onto a range of substrates. These different substrates interact with the CNT ink based on chemistry (wettability/contact angle, $\theta$) and topographic parameters (surface roughness, $r$). The contact angle $\theta$ is a measure of the balance of interfacial energies between liquid, solid, and vapor. Surface roughness, $r$, is defined here as the ratio of the actual area of the surface solid compared to the normally projected area. Both terms are assembled into a single apparent contact angle using Wenzel's law, $\cos(\theta^*) = r \cos\theta$, which predicts that roughness amplifies the effect of native surface wetting properties [53]. This parameter is listed below each image in Figure 4a-b, which are listed in order of increasing $\cos(\theta^*)$, and transition from $\cos(\theta^*) < 1$ for impermeable substrates in Figure 4a to $\cos(\theta^*) > 1$ for porous paper in Figure 4b.

A summary of the substrate parameters is shown Table 1. Contact angle measurements were taken from literature measurements of water on smooth substrates of the designated material, with references listed in Table 1, while roughness measurements were made using a laser scanning confocal microscope as described in Methods, Section IV. We note that the use of tabulated values for equilibrium contact angle is an approximation, and subsequent error may affect $\cos(\theta^*)$; however, this is required here due to the complexity of robustly measuring contact angle on a rough surface with a viscoelastic yield stress fluid.

Table 1: Substrate wetting parameters: contact angle $\theta$, roughness $r$, Wenzel wetting parameter $\cos(\theta^*)$. References are listed for contact angle values in the rightmost column.

| Substrate | $\theta\circ$ | $r$ | $\cos(\theta^*) = r\cos\theta$ | source for $\theta$ |
|---|---|---|---|---|
| Glossy paper | 68 | 1.3 | 0.47 | [91] |
| Hydrophobic paper | 80 | 3.0 | 0.52 | [92] |
| Vinyl ethylene film | 80 | 1.1 | 0.86 | [93] |
| Kapton film | 71 | 2.0 | 0.64 | [94] |
| Viscose filter | 80 | 5.5 | 0.95 | [92] |
| Polyethylene terephthalate | 70 | 3.0 | 1.02 | [95] |
| Chromatography paper | 0 | 7.2 | 7.21 | [91] |

A quasi-static equilibrium model was constructed to predict the width of lines printed onto impermeable substrates based on the wetting parameter $\cos(\theta^*)$ and the deposition rate through a balance of surface tension and gravity, as shown schematically in Figure S8. The deposition rate determines how much fluid (per unit depth) is extruded in a given area, having a height, $h$. The Bond number, $Bo = \rho g h^2/\gamma$, evaluates the relative influence of gravity and capillary forces on maintaining the droplet shape, where $g$ is the acceleration due to gravity, $\rho$ is the material mass density, $\gamma$ is the surface tension, and $h$ is the maximum height of the drop.

The model construction is shown here. The ink deposited (per depth) has an initial crosssectional area of

$$A_{cross} = Q_{ink}/V_{nozzle} = v\pi D_0^2/4 \tag{S6}$$

When this area is small such that the height of the printed trace is not substantially affected by gravity, $h < \lambda_c$ or $\sqrt{Bo} < 1$, the shape scales as $v \propto w^2$ or $w \propto \sqrt{v}$, with the constant of proportionality depending on the specific geometry. If the cross-section is simplified to be an isosceles triangular shape, then $A_{cross} = w^2 \tan\theta^*/4$, which is combined with the deposited fluid area to predict $w/D_0 = \sqrt{(\pi v/\tan\theta)} = a\sqrt{(v)}$. Other literature discusses finer resolution of droplet shape in Newtonian fluids [96], which we neglect here due to complex cross-sectional shapes that will arise from competing effects of yield stress, viscoelasticity, and compositiondependent surface tension of the CNT inks. In addition, we limit our analysis to cases in which the droplet is much larger than surface roughness so the substrate can be considered homogeneously rough. For very small feature sizes, if the roughness of the substrate approaches the size of the drop, gravity-driven spreading would be significantly altered [97]. In addition, this is similar to previous analysis for three-dimensional droplets on rough substrates [98].

When the cross-sectional area of the printed line is large, the cross-section becomes "puddlelike" and consists of a flattened central plateau with tapered edges, which varies as $w = b + cv$ through a similar geometric scaling analysis approximating the puddle as a trapezoid with central height $h$. The coefficients are $b = h/\tan\theta$ and $c = \pi D_0^2/4h$, where $h$ is the plateau height of the puddle, $h_c \approx \sqrt{2\gamma(1 - \cos\theta^*)/\rho g}$, marking the transition height between the puddle ($h < h_c$) and droplet ($h > h_c$) models, and ranges 1-2.2 mm here depending on $\theta$, and is similar to the capillary lengths for distilled water (2.6 mm) and for water with sodium deoxycholate (1.2 mm, with $\gamma \approx 44$ mN/m for 0.05 M sodium deoxycholate in water [99]).

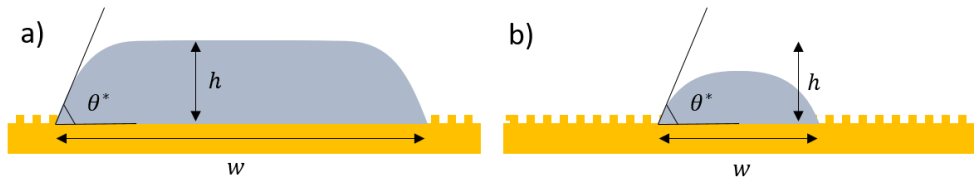

Figure S8: A diagram of (a) the puddle model and (b) the droplet model shows the assumed geometry where the puddle contacts the edge of the drop at a given contact angle and reaches an equilibrium shape with a maximum possible height $h_c$ set by a balance of gravity and surface tension/wetting effects.

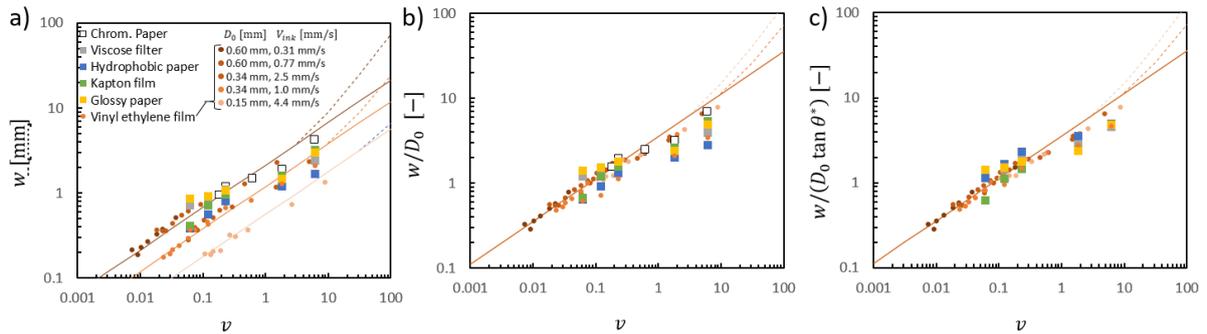

Figure S9: a) Measured data of printed line widths is compared to model predictions. (b) The vertical axis has been rescaled by the nozzle inner diameter, showing some collapse of the data. (c) The vertical axis has been rescaled by the nozzle inner diameter and the relevant substrate roughness parameter ($\tan\theta^*$) for each print, showing better collapse of data onto the model line for a range of process parameters and substrates.

### S1.IX. Monitoring drying time of printed CNT traces

While it was often apparent by eye when printed lines were mostly dry, we implemented an *in situ* resistance measurement on the printing stage to verify when the resistance had reached a stable level after solvent evaporated. As shown in SI Figure S10 for three cases, the drying time ranged between 1 and 21 minutes. The change in resistance was much sharper at the end for the plastic substrate, vinyl ethylene, than for the porous and absorbent chromatography paper.

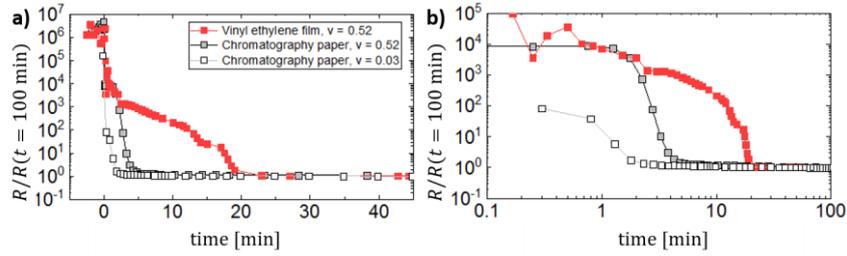

Figure S10: The resistance of printed lines was monitored after printing to determine the drying time. Colors are the same as in Figure 4.

### S1.X. Influence of the Substrate on Minimum Printing Resolution

Line edge roughness and substrate de-wetting influence the minimum achievable feature size and density, respectively. Due to the dewetting process, lines become discontinuous much sooner on nonwetting viscose filter paper compared to coated glossy paper, which is well-wetted by the ink (SI Figure S11a,b). Similarly, because line edges are smooth when printing onto Kapton film, lines can be printed very close together without shorting between them, increasing the attainable density of coils on a printed spiral (SI Figure S11c,d).

This was evaluated more precisely by measuring the edge roughness as the excess tortuosity, or

$$\text{roughness} = 1 - \frac{L_p}{L} \tag{S7}$$

where $L$ is the printed line length and $L_p$ is the length of its edge perimeter as viewed from microscopy images as shown in Figure 4, which has a minimum value of $L$. This roughness correlates strongly with both the roughness and effective contact angle of the substrate, as shown in SI Figure S12.

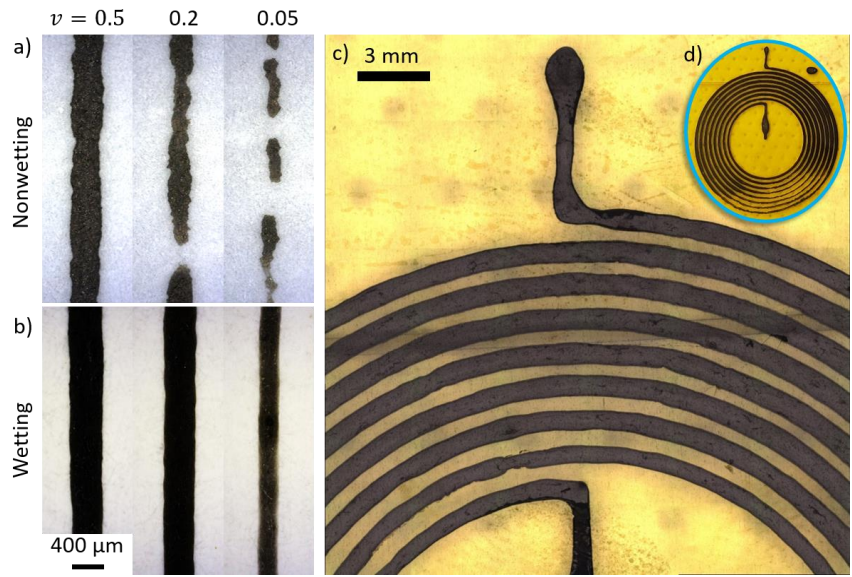

Figure S11: Printed straight lines are shown as a function of deposition rate, $v$, for (a) nonwetting (viscose filter) and (b) wetting (glossy paper) substrates. (c,d) Lines printed closely together are shown on Kapton film.

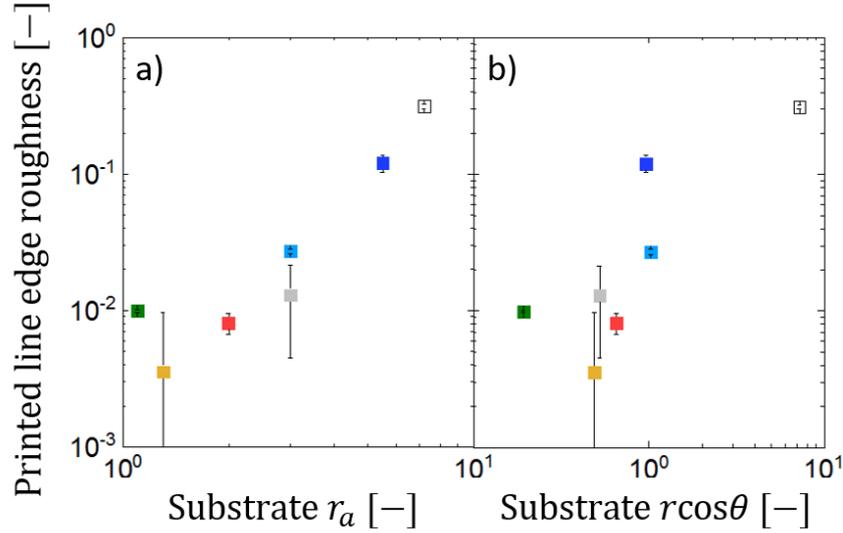

Figure S12: The roughness of printed lines is shown as a function of (a) the substrate roughness and (b) the substrate. Colors are the same as in Figure 4.

*S1.XI. Conductivity of CNT Lines for Repeated Bending Cycles*

The mean conductivity of our printed traces are constant within 2% for at least 1,000 bending cycles with a full inward crease (bending such that a crease is formed on the paper and the two inner faces on either side of the fold contact each other). However, for significantly less dense CNT traces with $v_{ink}/v_{nozzle} = 0.05$, the conductivity degrades linearly with the number of bend cycles up to a mean 28% increase in resistance over 1,000 bending cycles, as shown in SI Figure S13a. The folding cycles are performed manually, creating a full crease in lines between two electrical probes, as shown in SI Figure S13b.

These systematic changes are ascribed to adhesion limits between printed CNT lines and the substrate, and how that influences the material involved in the fold. For impermeable substrates, adhesion is much lower and so lines may partially delaminate, as shown in SI Figure S14, mechanically affecting a wider range of CNTs. When the substrate is porous, much less material is incorporated in the fold because it is more strongly adhered to the paper, which causes more acute destruction of the CNT network in the narrower region, and is more sensitive when less overall CNT material is deposited, denoted by low ink deposition, $v$.

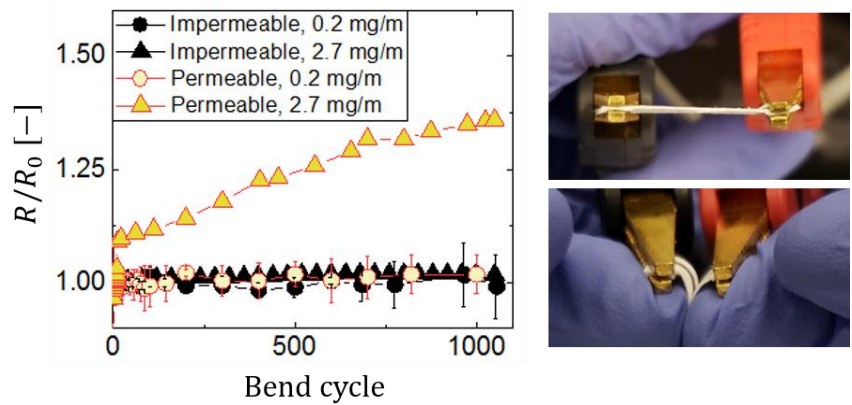

Figure S13: (a) The mean conductivity of our printed traces are constant within 2% for at least 1,000 bending cycles with a full inward crease. For significantly less dense CNT traces with $v_{ink}/v_{nozzle}$ = 0.05, the conductivity degrades linearly with bend cycles up to a mean 28% increase over 1,000 bending cycles, with an initial resistance $R_0 \approx 300-2100\Omega$. (b) The folding cycles are performed manually, creating a full crease in lines between two electrical probes.

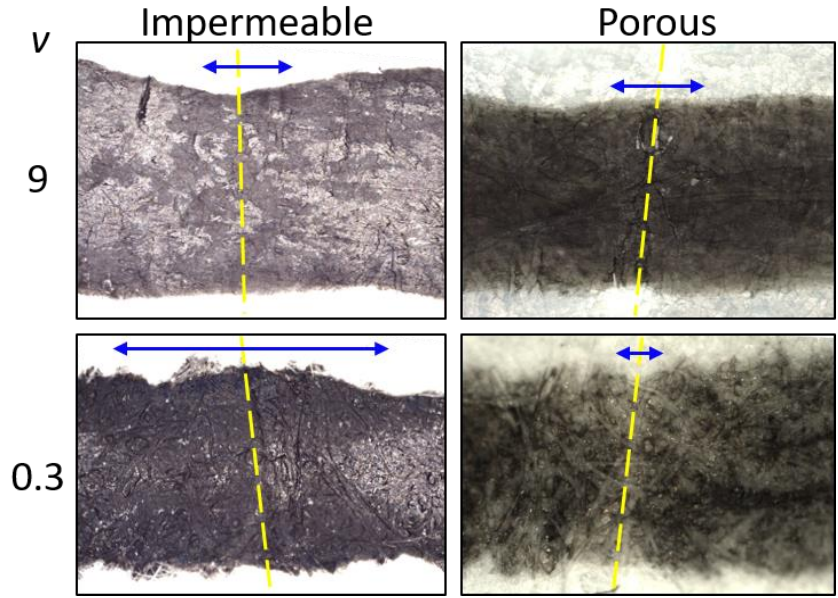

Figure S14: For impermeable substrates, adhesion of CNT traces is much lower and so lines may partially delaminate, mechanically affecting a wider (darker) spatial area of CNTs. When the substrate is porous, much less material is incorporated, which causes more acute destruction of the CNT network in the narrower (lighter) region, and this effect is amplified when less overall CNT material is deposited, denoted by low $v$ values. The lines shown for $v$ = 9 are approximately 3000 $\mu$m in width, and for $v$ = 0.3 are approximately 800 $\mu$m in width.

CNT lines showed stable resistance over repeated folding cycles, as shown in SI Figure S13a and in SI Section S1.XI. On both impermeable and permeable substrates, lines made with a linear density of $\rho A_c$ = 2.7 mg/m had a mean percentage change in resistance under 2%. This behavior continued for more sparse prints with $\rho A_c$ = 0.2 mg/m on impermeable substrates; however resistance increased and showed a greater change in resistance over 30% for lines printed onto porous substrates with the more sparse linear density of $\rho A_c$ = 0.2 mg/m.

Following the kirigami design work of Morikawa, et al.,[84] we created an ultrastretchable substrate capable of folding and extension, shown in SI Figure S15. CNTs deposited on this substrate showed DC conductivity with under 5% change for a single mechanical loading/unloading cycle up to 830% (maximum) strain, 100 cycles up to 400% strain, and under 1% change in resistance for 100 cycles at 200% strain. This extensibility can help one-time integration into devices (ex. for thermoforming), even at very high strains, as well as for repeated moderate (<500%) strain cycles. For comparison, human skin can generally stretch 50% [100], so wearable sensors would be well within this range.

In addition, these ultrastretchable CNT forms were able to deform in planar extension and in an out-of-plane bending deformation, in which they were raised vertically by a central tower composed of LEGO bricks, which are known for having precise dimensions [101].

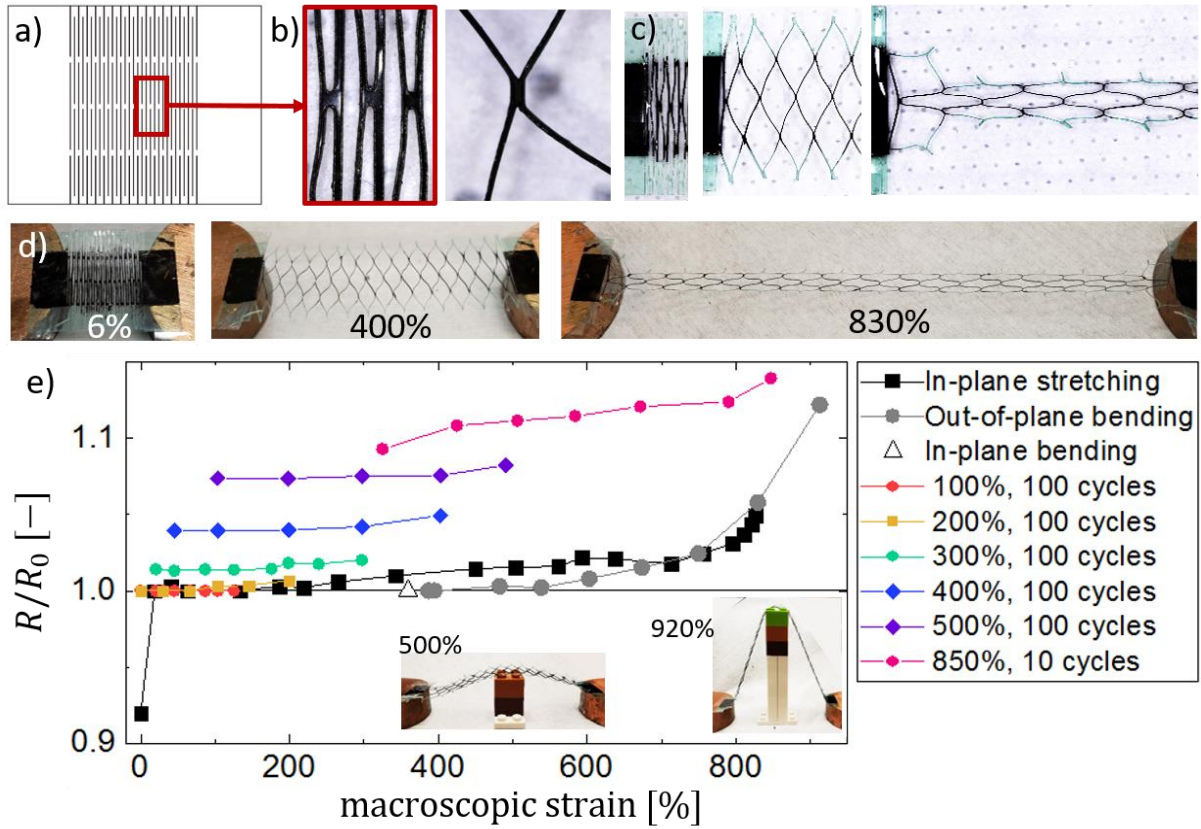

Figure S15: An ultrastretchable substrate was created with (a) a design of vertical cuts that (b) stretch and extend at a few internal joints to (c,d) extend elastically from unstrained to >900% strain. (e) The strain over time for the flexible substrate, for a single repetition and for sequential repeats of 100 cycles of extension up to the denoted maximum strain. Note that data for repeated cycles were performed in series, so 100 cycles were performed from 0 to 100% strain, and then 100 cycles were performed from 0 to 200% strain, and so on. Data for the in-plane stretching, out-of-plane bending, and in-plane bending only were done separately.

*S1.XII. Contact Resistance Measurements of CNT Traces*

The method of measuring the contact resistance consisted of measuring the conductivity of CNT traces for different printed lengths, using a configuration with a continuous CNT line and two probes, and a second configuration with a break in the CNT line to incorporate a controlled CNT-CNT contact point. From the data collected, which is shown in SI Figure S16, the slope is $R_m$. The intercept of the base measurement is $2R_{c,probe}$, and the intercept of the measurement including CNT-CNT contact is $2R_{c,probe} + R_c$.

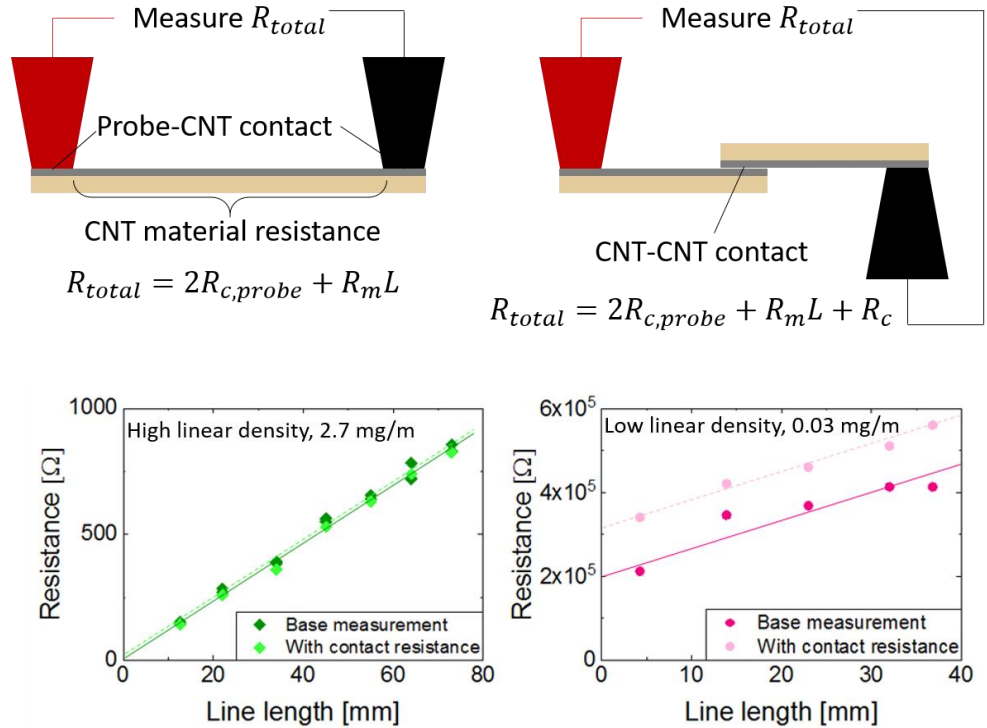

Figure S16: CNT material resistance was measured using contact of two probes with varied length between probes on the same substrate. By varying the length between probes, the probe-CNT contact was measured separately from the base material resistance. In a second test, a CNT-CNT contact was introduced, and the contact resistance was measured the same way by varying the total length of measurement for the same CNT line as measured in the first case. Example data sets are shown for CNT lines with high and low linear density.

The variation of resistance and contact resistance was evaluated over a series of cycles of applied pressure for CNT traces printed onto glossy and chromatography paper, as shown in SI Figure S17. In displacement-controlled tests, pressure normal to the substrate was applied, resulting in cycles of high pressure alternating with loss of contact with the substrate (SI Figure S17a-b). The conductance (defined as the inverse of the resistance) was monitored simultaneously (SI Figure S17c,e) and the ratio of the contact resistance to the trace resistance over 1 mm was calculated to show the cyclic behavior (SI Figure S17d,f). While the results for the glossy paper were repeatable, the behavior of the chromatography paper drifted over time before reaching a more steady behavior, which we ascribe to plastic deformation of the chromatography paper itself (data not shown). After several cycles, the comparison between substrates continued to show that CNT traces printed onto glossy paper had a much sharper change in contact resistance with applied pressure, while for those printed onto chromatography paper, the response became nearly independent of pressure. This behavior could have strong implications for the design of CNT contact-pressure buttons that are either sensitive or insensitive to applied pressure.

We note that these results are dependent on the behavior of the substrate so that similar papers with different thicknesses or moduli would be expected to behave differently, as well as papers with different mechanisms preventing separation. The glossy paper we used is both thinner (around 120 μm) and denser than the chromatography paper (around 150 μm in thickness).

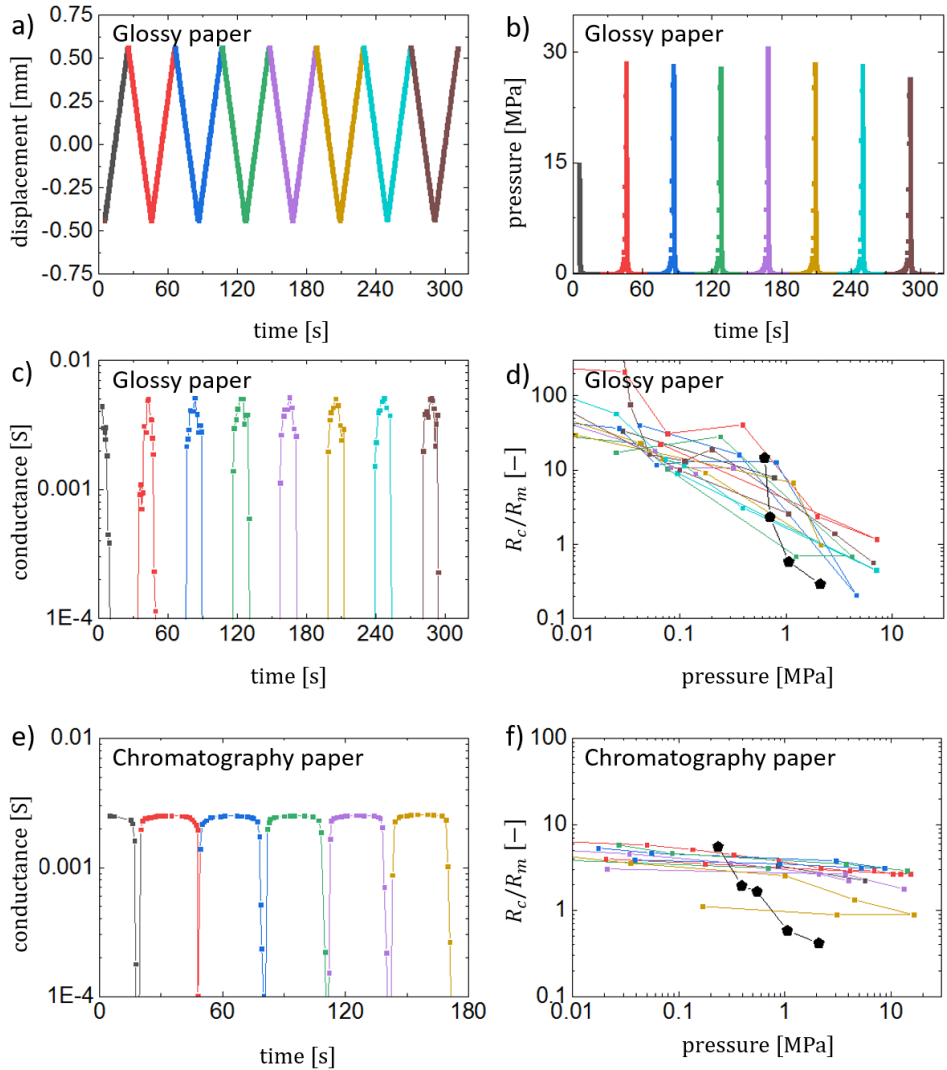

Figure S17: The contact point between two CNT traces printed onto paper was loaded to apply cyclic compressive forces. (a) The cycles were controlled by displacement, which (b) generated an applied pressure. (c,e) The conductance (defined as the inverse of resistance) was monitored over the same cycles and (d,f) the contact resistance was compared to pressure over several cycles. The data with black circles in (d,f) here is reproduced for the same substrate and conditions as shown in Figure 6a,b.

*S1.XIII. Fluid Sensor Used For Droplet Impact*

A two-line CNT-based printed sensor was used to register the impact of a small droplet of distilled water (SI Figure S18). The sensor exhibits a sensitivity to the volume of the droplet through the amplitude of the initial spike in response and the area under the full curve. The capacitive sensor can register addition of multiple droplets in quick succession (red arrows) and is still sensitive to the total volume, where the response after 60 seconds is equal for equal volumes deposited as separate drops or all at once. The experimental approach deposited droplets onto the CNT-based printed sensor using a calibrated micropipette.

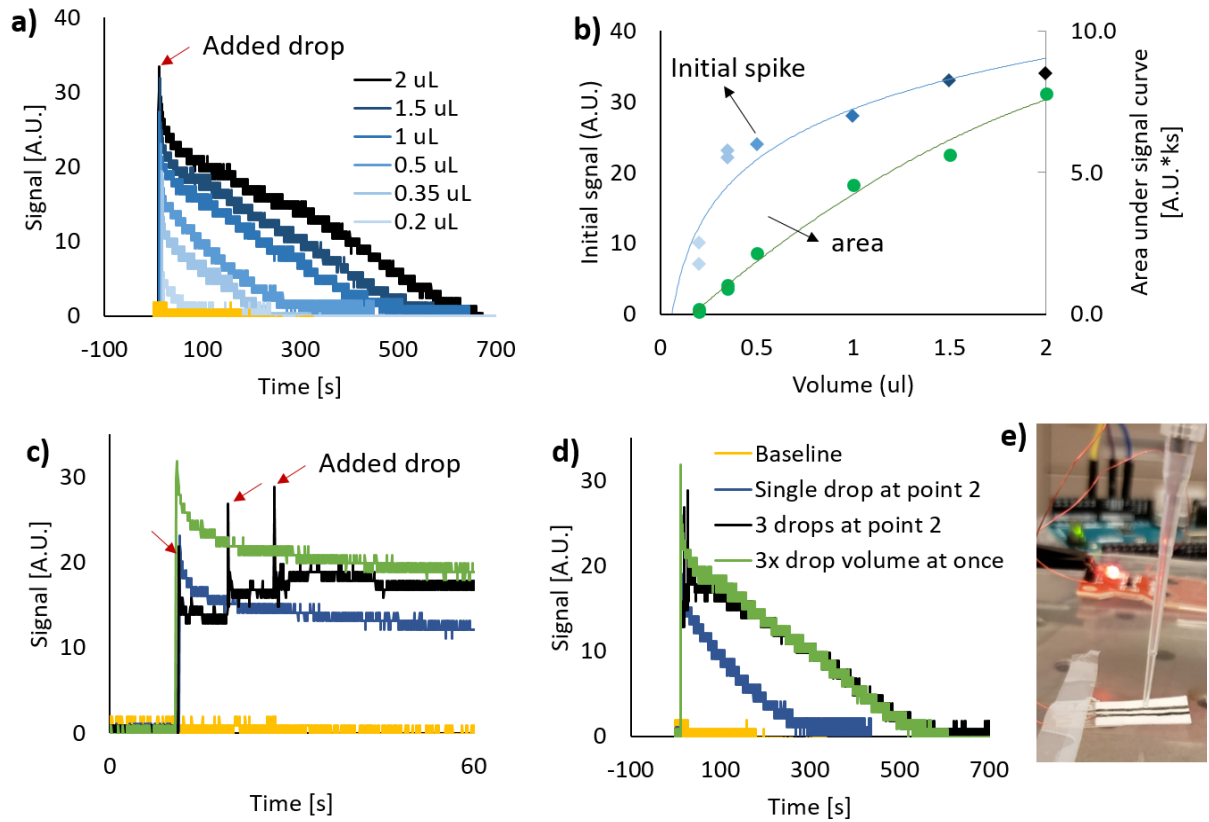

Figure S18: (a) Voltage signal recorded upon measuring the impact of a small droplet of distilled water on a twoline printed CNT capacitive sensor. (b) The sensor is sensitive to the volume of the droplet through the initial spike in response and the area under the full curve. (c,d) The sensor can register addition of multiple droplets in quick succession (red arrows) and is still sensitive to the total volume, where the response after 60 seconds is equal for equal volumes deposited as separate drops or all at once. Here, each drop is a constant volume of 0.5 $\mu$L. (e) The experimental setup is shown, where droplets are deposited by a calibrated micropipette.